\documentclass[journal, 12pt, draftclsnofoot, onecolumn]{IEEEtran} 
\usepackage{amsmath,amsfonts}
\usepackage{algorithmic}
\usepackage{algorithm}
\usepackage{array}
\usepackage[caption=false,font=normalsize,labelfont=sf,textfont=sf]{subfig}
\usepackage{textcomp}
\usepackage{stfloats}
\usepackage{url}
\usepackage{verbatim}
\usepackage{graphicx}
\usepackage{cite}
\usepackage{amssymb} 
\usepackage{upgreek}
\usepackage{booktabs}
\usepackage{multirow}
\usepackage{makecell}
\begin{document}

\title{Enhancing Feature Extraction for Indoor Fingerprint Localization Using Diversified Data}

\author{Jiyu Jiao, Xiaojun Wang, Chenlin He
\thanks{This work was supported in part by the National Key R$\&$D Program of China under Grant 2022YFC38010000 and in part by the Key Research $\&$ Developement Plan of Jiangsu Province under Grant BE2020084-2 and in part by the Fundamental Research Funds for the Central Universities under Grant 2242022k60001. \textit{(Corresponding author: Xiaojun Wang)}}
\thanks{Jiyu Jiao and Chenlin He are with the School of Information science and Engineering, Southeast University, Nanjing 214135, China (e-mail: jiyu$\_$jiao@seu.edu.cn;	clhe@seu.edu.cn).}
\thanks{Xiaojun Wang is with National Mobile Communications Research Laboratory, Frontiers Science Center for Mobile Information Communication and Security, School of Information science and Engineering, Southeast University, Nanjing 211100, China and Purple Mountain Laboratories, Nanjing 211111, China (e-mail: wxj@seu.edu.cn).}
}



\maketitle

\begin{abstract}
Given the rapid advancements in wireless communication and terminal devices, high-speed and convenient WiFi has permeated various aspects of people's lives, and attention has been drawn to the location services that WiFi can provide. Fingerprint-based methods, as an excellent approach for localization, have gradually become a hot research topic. However, in practical localization, fingerprint features of traditional methods suffer from low reliability and lacking robustness in complex indoor environments. To overcome these limitations, this paper proposes a innovative feature extraction-enhanced intelligent localization scheme named Secci, based on diversified channel state information (CSI). By modifying the device driver, diversified CSI data are extracted and transformed into RGB CSI images, which serve as input to a deep convolutional neural network (DCNN) with SE attention mechanism-assisted training in the offline stage. Employing a greedy probabilistic approach, rapid prediction of the estimated location is performed in the online stage using test RGB CSI images. The Secci system is implemented using off-the-shelf WiFi devices, and comprehensive experiments are carried out in two representative indoor environments to showcase the superior performance of Secci compared to four existing algorithms. 

\end{abstract}


\section{Introduction}
\IEEEPARstart{W}{ith}  the advancement of Internet of Things (IoT) technology and the increasing ubiquity of smart mobile devices, there is a growing demand for location-based services to cater to the needs of smart living\cite{zhao2022high}, including applications such as rescue localization, trajectory tracking, and robot navigation\cite{roy2021survey}. In the realm of indoor localization, WiFi-based positioning technology has gained attention due to its wide availability in indoor environments. However, the accuracy and reliability of indoor localization remain critical issues that require further research and improvement, owing to challenges posed by complex indoor environments, multipath effects, and obstructions. Fingerprint-based methods have emerged as an excellent approach to address these challenges and have become a research hotspot\cite{chen2021adaptive}. The fundamental idea is to collect signal features from multiple discrete points within a localization area and combine them to form a fingerprint database. During localization, matching is performed with the fingerprints in the database, selecting the most similar fingerprint points to estimate the location based on certain rules.

To achieve precise indoor localization, various technologies have been proposed, such as WiFi\cite{zou2017winips}, ultrasonic\cite{ward1997new}, radio frequency identification (RFID)\cite{bernardini2020robot}, Zigbee\cite{angrisani2017analysis}, Bluetooth\cite{lee2018indoor}, ultra-wideband\cite{yang2021novel}, and infrared\cite{tao2014multiperson}, aiming to overcome the limitations of traditional wireless localization techniques, including drawbacks such as high electromagnetic radiation interference, high deployment costs, and low positioning accuracy. WiFi, in particular, has gained significant attention due to its reliability, extensive coverage, and ability to fill the gaps left by satellite-based positioning systems. Many fingerprint-based localization approaches have been developed based on WiFi\cite{bahl2000radar}. Commonly used fingerprints in WiFi-based methods include Received Signal Strength Indicator (RSSI)\cite{zou2015robust}, Channel State Information (CSI)\cite{long2022adaptive}, and combinations of WiFi with magnetic fields or Bluetooth\cite{shao2018indoor}\cite{luo2019indoor}. To improve localization accuracy, the advancements in artificial intelligence (AI) have facilitated the use of deep learning (DL), which possesses powerful learning capabilities, to extract RSSI features for localization\cite{mazlan2022fast}. However, RSSI provides coarse-grained information, limiting the localization accuracy\cite{sobehy2020csi}.

Channel State Information (CSI) is a new measurement metric that provides fine-grained physical layer information at the subcarrier level, encompassing more detailed and diverse physical layer characteristics of signal propagation. As a result, it has found applications in gesture recognition, location tracking\cite{shi2018accurate}, and indoor positioning. Researchers have also focused on addressing indoor positioning challenges using CSI-based approaches\cite{xiao2012fifs, gjengset2014phaser, zhu2022intelligent}. Some network interface cards (NIC) compliant with the IEEE 802.11n standard can represent received signals in the form of CSI, providing users with the amplitude and phase values of individual subcarriers. However, several challenges need to be addressed. First, in non-line-of-sight(NLOS) indoor environments, fluctuations in subcarrier amplitudes at different sampling instances diminish the effectiveness of CSI as a positioning feature. Second, the high-dimensional nature of CSI data incurs significant positioning time overhead. Third, there is a need to model the relationship between CSI-based positioning features and position coordinates.

In this paper, we propose the Secci system, which utilizes image processing to address the aforementioned issues. CSI encompasses frequency responses of multiple subcarriers and allows for feature extraction of phase and amplitude values. Due to the stability of phase differentials, amplitudes, and phase data, estimated angle of arrival (AoA) and average amplitudesare relatively more stable. Even when the signal is obstructed, they can remain relatively stable. Therefore, for complex indoor environments, the combination of average amplitudes, AoA estimation, and phase information complements each other and exhibits high robustness. Instead of employing geometry-based angle estimation techniques, we utilize a DCNN augmented with an attention mechanism to extract features from diverse data. To enhance the network's ability to extract discriminative features, the SE attention mechanism is employed to quantify the importance of different channels, providing a reliable approach. We compute 270 diverse data points from each packet captured by the Intel 5300 NIC and construct 33 images of size 90$\times$90 using the diverse data from 2970 received packets as input to train the Secci system's weights. Additionally, we propose a greedy probabilistic approach to compute the probability of estimated positions.

The paper's main contributions can be summarized as follows:

\begin{itemize}
	\setlength{\parskip}{0pt} 
	
	\item {We employ average amplitude values, estimated AoA values, and phase values of CSI data, transforming them into RGB CSI images as fingerprints for feature learning. This straightforward data transformation avoids complex preprocessing, leading to low computational complexity. We provide theoretical and experimental validation of the feasibility of utilizing these diverse features for indoor localization. Furthermore, we demonstrate that these three CSI features complement each other, leading to robust localization performance in indoor environments.}
	\item {We design the Secci system, a DCNN network assisted by an attention mechanism, which enables end-to-end estimation of the mobile device's location. We develop an offline training algorithm for CSI images in the system, and propose a greedy probabilistic method for online location estimation.}
	\item {The Secci system is implemented using an Intel 5300 NIC and evaluated in two typical indoor environments. Experimental results show that the Secci system surpasses four existing schemes. Furthermore, a comprehensive analysis is conducted to examine the factors influencing the performance of the Secci system.}
\end{itemize}

The remaining sections of this paper are organized as follows. Section II presents the preliminaries. Section III describes the Secci system. Section IV presents the experimental results and performance analysis. Section V discusses related work.  Finally, Section VI concludes this paper.

\section{Preliminaries}
\label{s2}
\subsection{CSI}
\label{CSI}
Orthogonal Frequency Division Multiplexing (OFDM) is a modulation technique that employs frequency division multiplexing by partitioning a frequency channel into multiple independent and orthogonal subchannels. It enables the parallel transmission of high-speed serial data and provides strong interference resistance. In comparison to RSS, CSI provides a more comprehensive reflection of various factors, including amplitude attenuation, phase offset, and time delay. Additionally, CSI captures information about individual subcarriers, allowing for a better characterization of signal propagation paths.
\begin{figure}[!ht]
	\centering
	\includegraphics[width=0.4\textwidth]{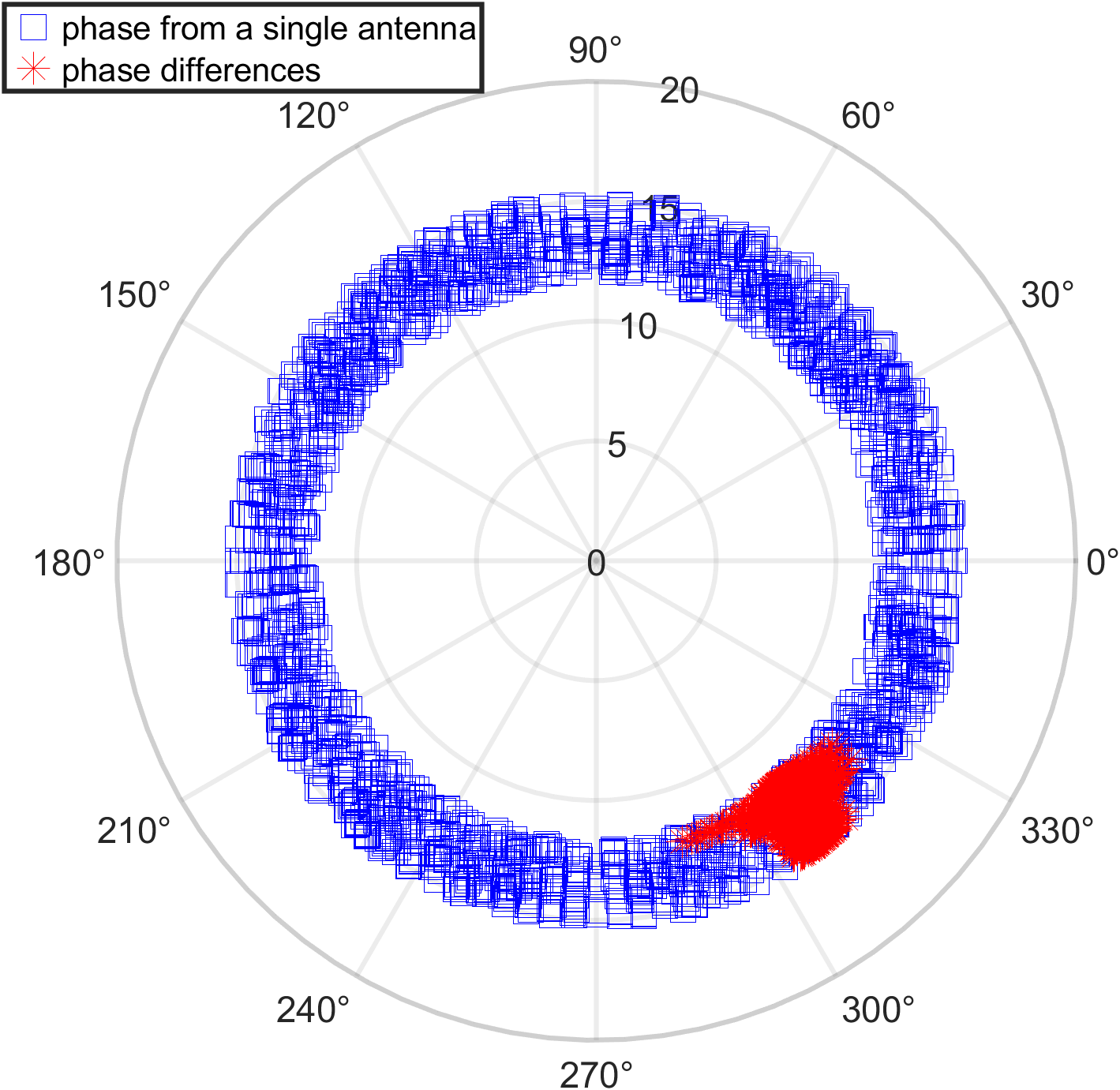}
	\caption{A comparison between phase differences (marked as red stars) and phases from a single antenna (marked as blue squares) of subcarrier 1 in the polar coordinate plot for 2000 back-to-back packets.}
	\label{fig1}
\end{figure}

CSI represents the impact of environmental factors on the signal propagation process and provides a detailed characterization of the channel characteristics between the transmitter and the receiver. In the frequency domain, this channel can be modeled as follows:
\begin{equation}
	\vec{R}  = CSI \cdot \vec{T} + \vec{N} ,
	\label{eq1}
\end{equation}
Where $\vec{R}$ represents the signal vector at the receiver, $\vec{T}$ represents the signal vector at the transmitter, and $\vec{N}$ represents the noise.

Furthermore, multipath effects can also be characterized by wireless CFR in terms of amplitude-frequency and phase-frequency characteristics. If the bandwidth is infinite, CFR and CIR are Fourier transforms of each other. The CSI for the $i$-th subcarrier (CSI$_i$) can be expressed as follows: 
\begin{equation}
	CSI_i  = \mathcal{I}_i + j\mathcal{Q}_i = \vert CSI_i \vert \exp({j\angle CSI_i }),
	\label{eq2}
\end{equation}
Where $\mathcal{I}_i$ represents the in-phase component and $\mathcal{Q}_i$ represents the quadrature component, $\vert CSI_i \vert$ and $\angle CSI_i$ represents the amplitude and phase of the $i$-th subcarrier, respectively.

\subsection{Phase Difference Information}
\label{Phase Difference}
Next we explore the importance of phase difference information and its conversion into estimated AOA information. The CSI phase data obtained from the Intel 5300 NIC is typically subject to randomness caused by unsynchronized time and frequency between the transmitter and receiver NICs, as well as environmental noise. To address this, two effective methods for CSI phase calibration have been proposed. The first method involves a linear transformation of coarse phase values \cite{qian2014pads}, \cite{wang2016csi}. The second method utilizes the phase difference between adjacent antennas \cite{wu2015phaseu}. However, the approach in \cite{wu2015phaseu} measures the average phase difference and removes it, and the proposed scheme is primarily designed for real-time line-of-sight (LOS) identification in 2.4 GHz WiFi, rather than an indoor localization solution. Furthermore, we opted to conduct data collection in the 5GHz frequency band for WiFi. This decision is supported by the findings in \cite{wang2017biloc}, which demonstrated that the performance of CSI-based indoor localization is superior in the 5GHz frequency band compared to the 2.4 GHz band.

In order to demonstrate the stability of phase difference between two antennas for consecutively received packets, we first model the measured phase of subcarrier $ i $ as follows\cite{xie2015precise}\cite{27}:
\begin{equation}
	\angle \widehat{CSI}_i  = \angle CSI_i + (\lambda_p + \lambda_s)m_i + \lambda_c + \beta + Z ,
	\label{eq388}
\end{equation}
where $\angle CSI_i$ is the true phase, $m_i$ represents the subcarrier index of subcarrier $i$, $Z$ represents the measurement noise, $\beta$ denotes the initial phase offset introduced by the phase-locked loop (PLL), and $\lambda_c$, $\lambda_s$, and $\lambda_p$ represent the phase errors resulting from central frequency offset (CFO), sampling frequency offset (SFO), and packet boundary detection (PBD) respectively \cite{27}\cite{28}, expressed as
\begin{equation}
	\begin{cases}
		\lambda_p =  \dfrac{2\pi \Delta t}{N} \\
		\lambda_s =  {\dfrac{2\pi n T_s}{T_u}}(\dfrac{T^{\prime} - T}{T}) \\
		\lambda_c = 2\pi n\Delta f T_s  ,
	\end{cases}\\
	\label{eq4}
\end{equation}
where $ \Delta t $ denotes the packet boundary detection delay, $ N $ denotes the FFT size, $ T$ and $ T^{\prime} $ denote the sampling periods of the transmitter and receiver, respectively, $ T_u $ is the length of the data symbol, $ T_s $ is the length of the data symbol plus the guard interval, $ \Delta f $ is the difference in center frequencies between the transmitter and receiver, and $ n $ denotes the sampling time offset for the current packet. The exact values of $ \Delta t $, $\beta  $, $ n $, $T^{\prime} - T  $, and $\Delta f$ in equations (4) and (5) are unknown and cannot be obtained. Moreover, $ \lambda_p $, $ \lambda_s $, and $ \lambda_c $ vary among packets with different $ \Delta t $'s and $ n $'s. Therefore, it is not possible to derive the true phase $ \angle CSI_i $ from the measured phase values.

Considering the synchronized clock and down-converter frequency shared by the antennas of the Intel 5300 NIC, errors resulting from frequency differences, sampling periods, and packet detection delay can be disregarded. Thus, we can approximate the measured phase difference between any two antennas for subcarrier $ i $ as follows:
\begin{equation}
	\Delta \angle \widehat{CSI}_i  = \Delta \angle CSI_i + \Delta \beta + \Delta Z ,
	\label{eq5}
\end{equation}
where $ \Delta \beta $ denotes the unknown difference in phase offsets, which is actually a constant \cite{gjengset2014phaser}, and $ \Delta Z $ is the noise difference. $ \Delta CSI_i $ is the true phase difference of subcarrier $ i $. Due to the absence of $ \Delta t $ , $ \Delta f $ and $ n $ in the above equation, $ \Delta \angle \widehat{CSI}_i $ is stable for different packets.

Under high SNR conditions, the measured phase difference of subcarrier $ i $ also conforms to $ \mathcal{N}(\Delta \beta, 2\sigma^2(1+1/{\vert CSI_0 \vert}^2)) $ distribution. This distribution arises from the independent phase responses\cite{wang2017biloc}. Although the variance is larger, the errors caused by time and frequency differences are eliminated, resulting in more stable measurement results, as shown in Fig.~\ref{fig1}. Fig.~\ref{fig1} depicts the phase differences (represented by red stars) and the phases from a single antenna (represented by blue squares) of subcarrier 1 in 2000 consecutive received packets in a polar coordinate system. It can be observed that the phase values from the single antenna are uniformly distributed between 0 and 360°, while the phase differences of the same subcarrier from two antennas concentrate within a sector ranging from 295° to 322°.

\begin{figure}[!ht]
	\centering
	\includegraphics[width=0.37\textwidth]{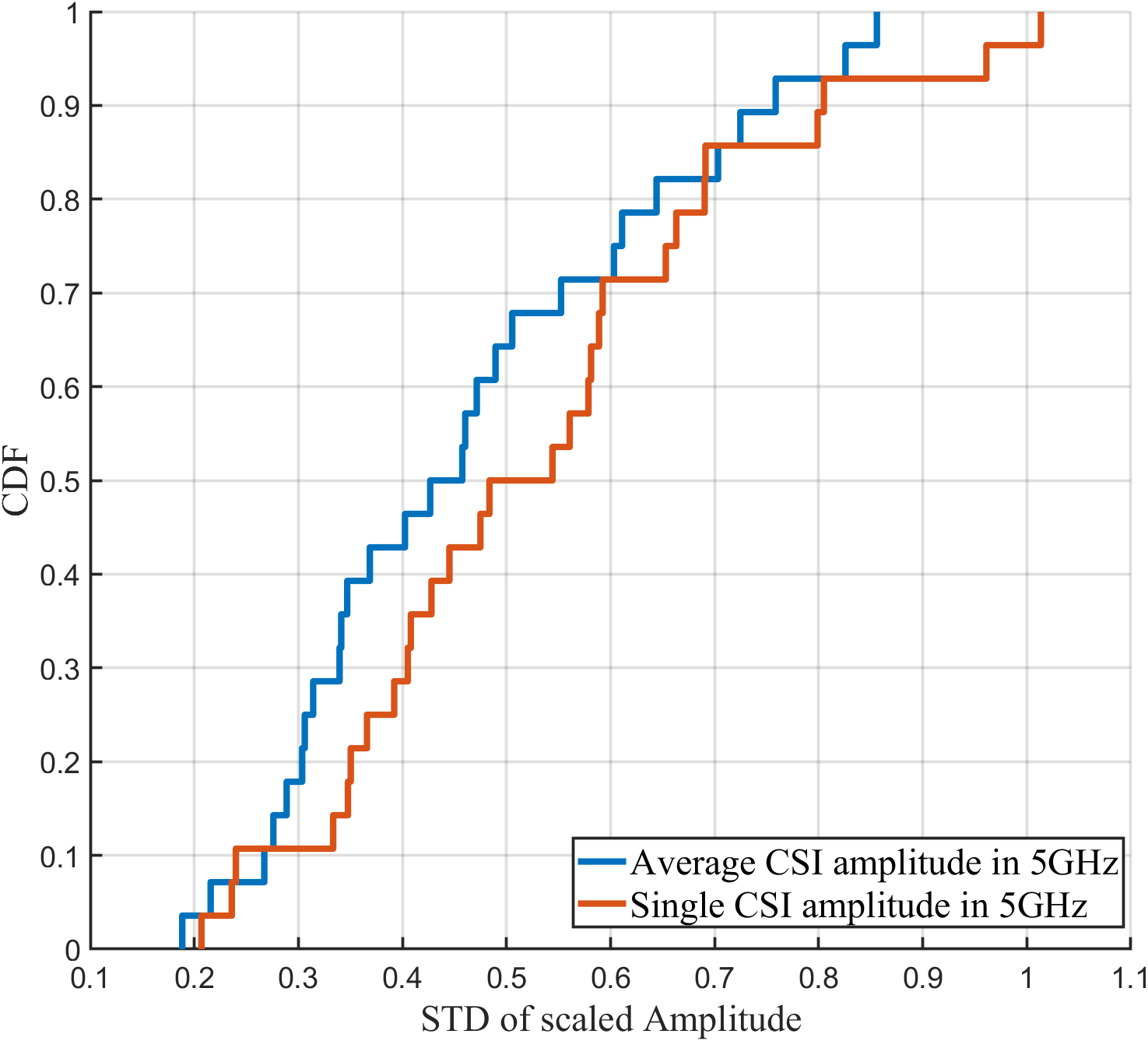}
	\caption{The CDF of the standard deviations (STD) for scaled average CSI amplitude and single CSI amplitude is calculated for subcarrier 15 in the 5GHz OFDM channel at 28 different coordinates.}
	\label{STD_CDF}
\end{figure}

Once the phase difference is obtained, the estimation of the Angle of Arrival (AoA) can be calculated using the following procedure:
\begin{equation}
	\theta_i = \arccos (\dfrac{\Delta \angle \widehat{CSI}_i \lambda}{2\pi d}) ,
	\label{eq6}
\end{equation}
where $ \lambda $ is the wavelength and $ d $ is the distance between two adjacent antennas. The measured phase difference is relatively stable, which leads to a more stable estimation of the AoA. This stability makes it suitable for precise indoor localization. In this paper, we set $ d = 0.5\lambda $, and the estimated AoA is within the range of $\left[ 0, \pi \right] $.

\subsection{Average Amplitude Information}
\label{Average Amplitude}
In this section, we demonstrate the stability of the average CSI amplitude between two antennas in continuously received packets. When a strong LOS component is present, the amplitude response follows a Rician distribution. The amplitude response's probability distribution function (PDF) is defined as follows:
\begin{equation}
	\label{eq7}
	\begin{split}
		\operatorname{\textit{f}} (\left| CSI_i \right| )&=\dfrac{\left| CSI_i \right|}{\sigma^2}\exp \left(-\dfrac{ {\left| CSI_i \right|}^2 + {\left| CSI_0 \right|}^2 }{2 \sigma^2} \right)\\&\times I_0 \left( \dfrac{ {\left| CSI_i \right|} \cdot{\left| CSI_0 \right|} }{ \sigma^2 } \right) ,
	\end{split}
\end{equation}
Where $ I_0(\cdot) $ represents the zeroth order modified Bessel function of the first kind, and $ \left| CSI_0 \right|  $ refers to the amplitude response without noise. In the high signal-to-noise ratio (SNR) scenario, the PDF $ \operatorname{\textit{f}}(\left| CSI_i \right|) $ tends to converge to a Gaussian distribution $ \mathcal{N}(\sqrt{{\left| CSI_0 \right|}^2+ \sigma^2}, \sigma^2) $\cite{akbar2015amplitude}. When the CSI values of two antennas are independently and identically distributed (i.i.d.), the average CSI amplitude also follows $ \mathcal{N}(\sqrt{{\left| CSI_0 \right|}^2+ \sigma^2}, \sigma^2/2) $ distribution with a smaller variance.

As shown in Fig.~\ref{STD_CDF}, we observe that the average CSI amplitude exhibits stronger stability in continuously collected packets at arbitrary positions. The standard deviation (STD) of the average CSI amplitude is below 0.5 for 65$ \% $ of the positions, compared to 50$ \% $ for the single antenna CSI amplitude under the same conditions. This demonstrates that averaging the CSI amplitudes over two antennas can enhance stability\cite{kleisouris2008impact}. Furthermore, Fig.~\ref{STD_CDF} also indicates that the overall stability of the CSI amplitude values is not particularly outstanding. To achieve accurate localization, we consider estimated AoA information and phase information in Secci, as they complement each other and serve as features for DL.

\begin{figure*}[!ht]
	\centering
	\subfloat[\label{fig:arm1} {\small \textit{Channel R at location 1}}]{\includegraphics[width=.3\linewidth]{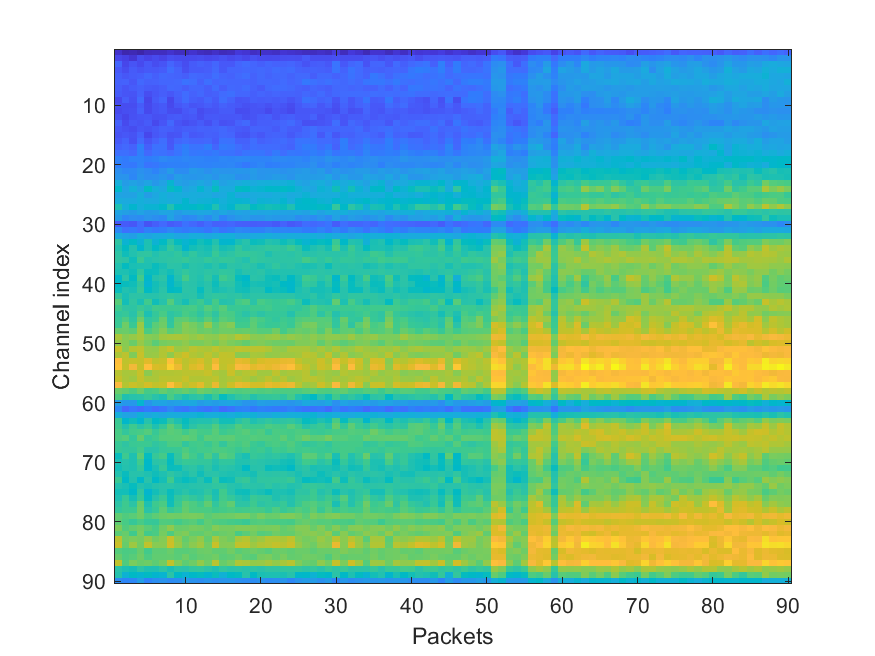}}
	\subfloat[\label{fig:arm2}{\small \textit{Channel R at location 2}}]{\includegraphics[width=.3\linewidth]{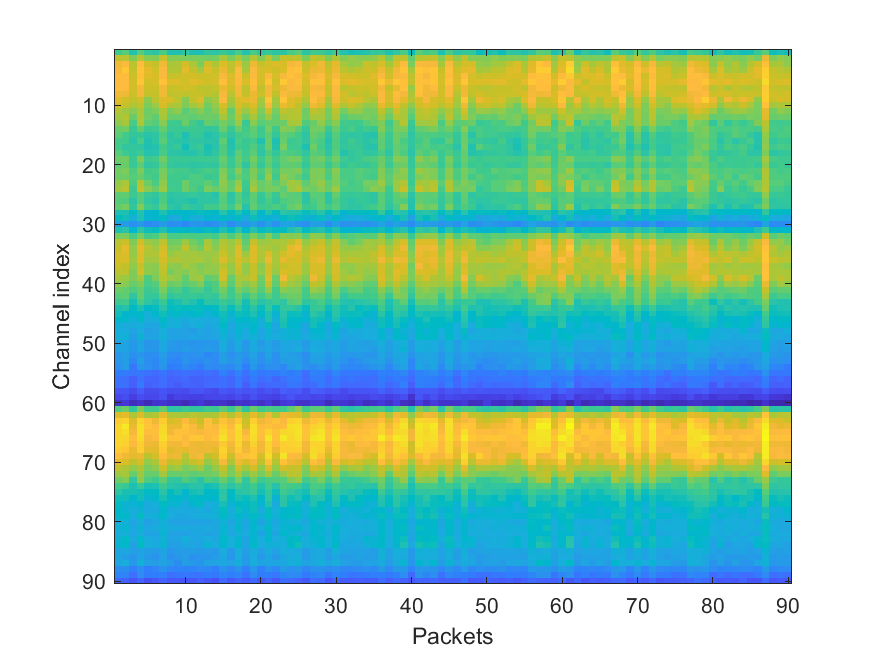}}
	\subfloat[\label{fig:arm3}{\small \textit{Channel R at location 3}}]{\includegraphics[width=.3\linewidth]{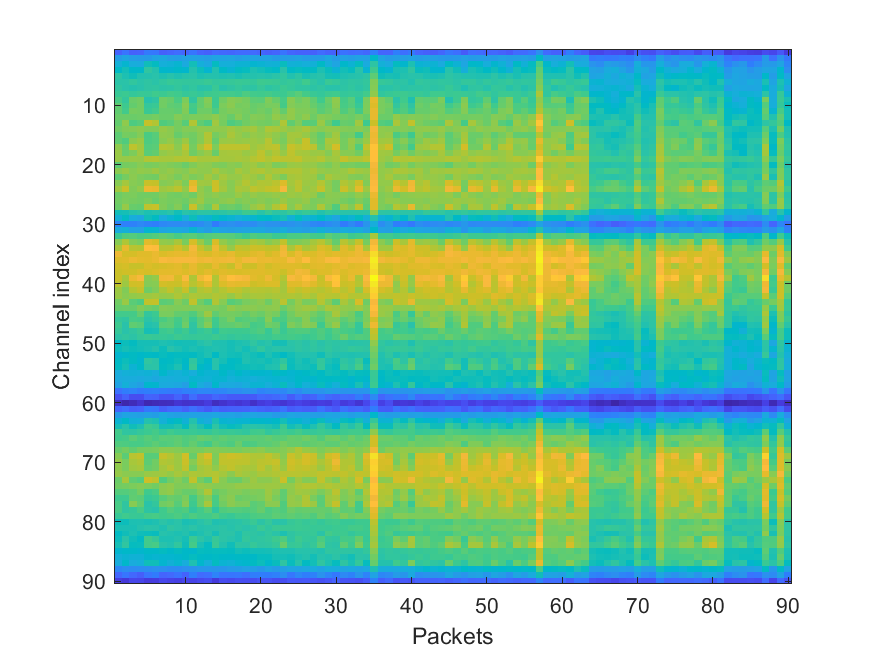}}\\
	\vspace{-5px}
	\subfloat[\label{fig:arm11}{\small \textit{Channel G at location 1}}]{\includegraphics[width=.3\linewidth]{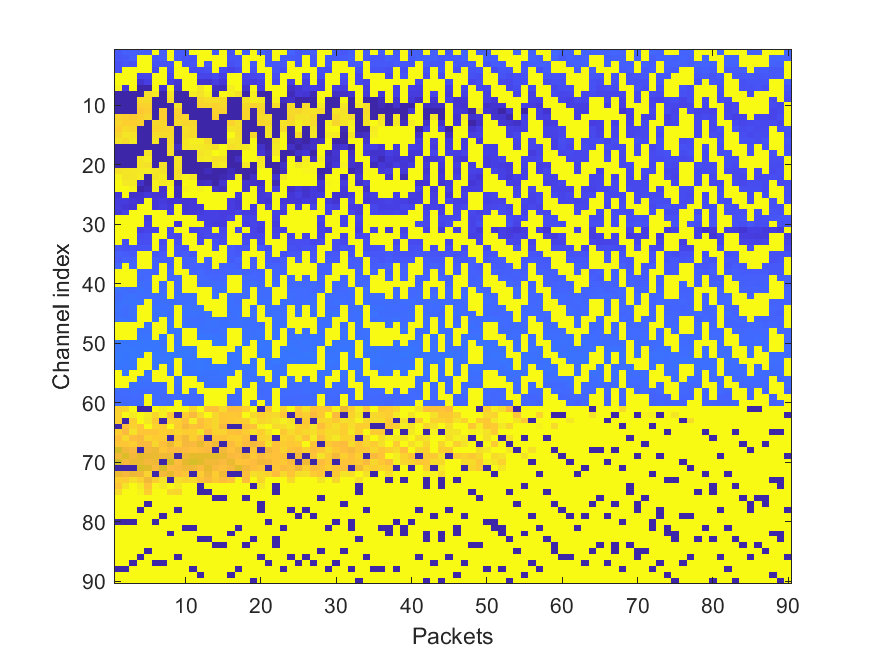}}
	\subfloat[\label{fig:arm22} {\small \textit{Channel G at location 2}}]{\includegraphics[width=.3\linewidth]{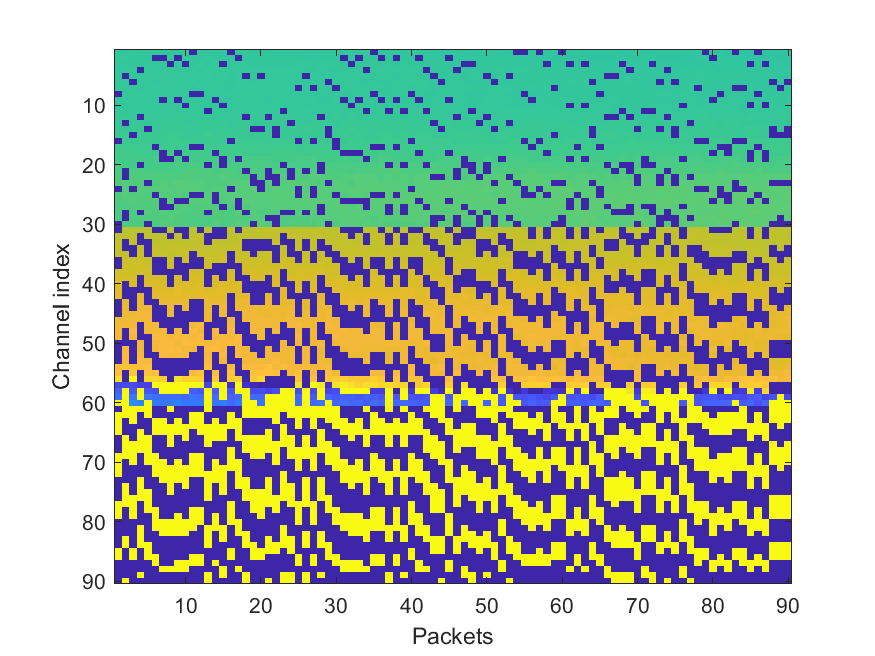}}
	\subfloat[\label{fig:arm33} {\small \textit{Channel G at location 3}}]{\includegraphics[width=.3\linewidth]{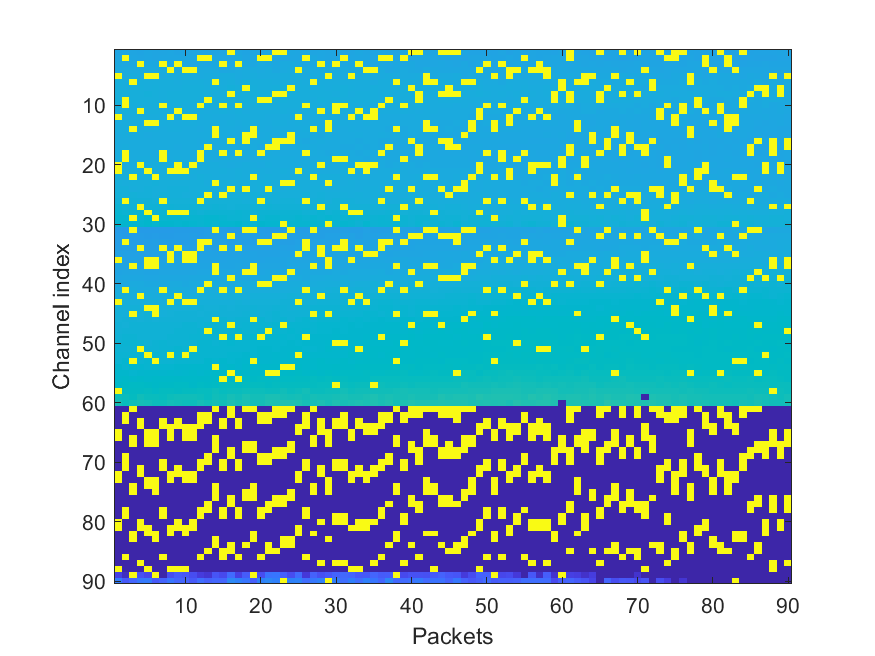}}\\%
	\vspace{-5px}
	\subfloat[\label{fig:arm111} {\small \textit{Channel G at location 1}}]{\includegraphics[width=.3\linewidth]{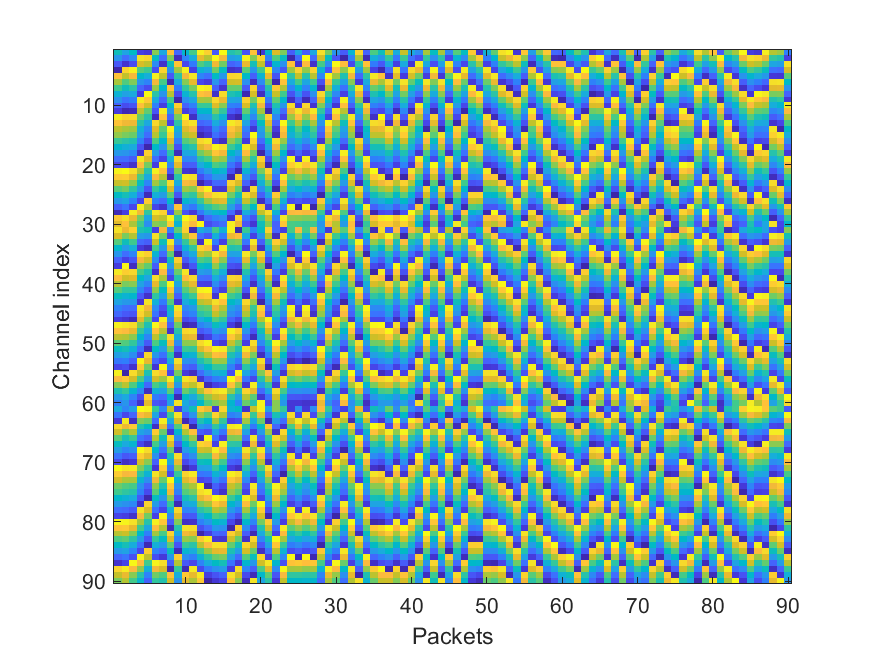}}
	\subfloat[\label{fig:arm222} {\small \textit{Channel G at location 2}}]{\includegraphics[width=.3\linewidth]{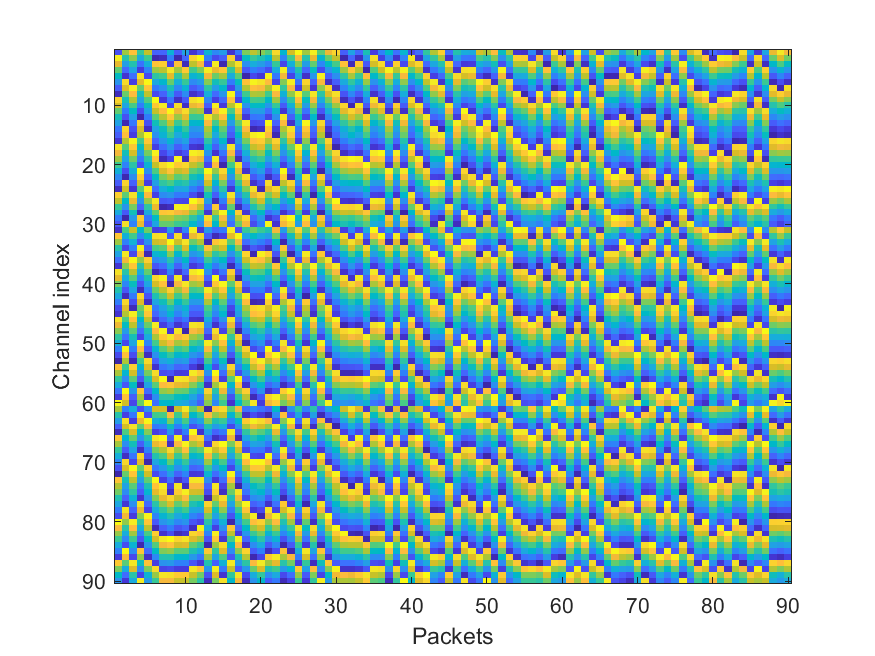}}
	\subfloat[\label{fig:arm333} {\small \textit{Channel G at location 3}}]{\includegraphics[width=.3\linewidth]{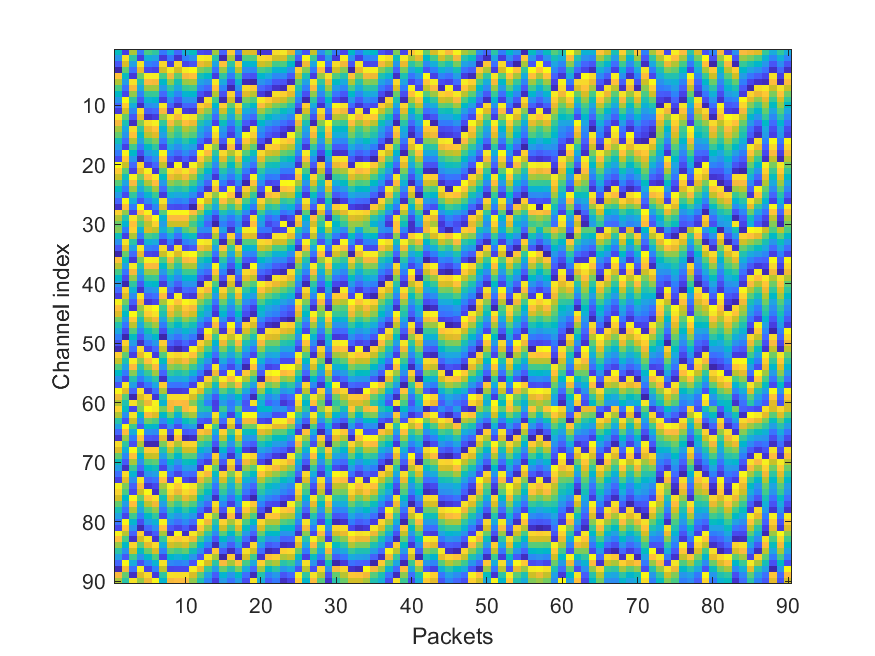}}%
	\caption{CSI images with three channels (average amplitude, estimated AOAs, phase) at three different locations.}
	\label{fig2}
\end{figure*}

\subsection{Image Construction}
\label{Image Construction}

The CSI data collected from Intel 5300 NIC includes the values of 30 subcarriers for each of the 3 antennas. We then use these values to calculate 30 phase values for each antenna separately: phase differences between antennas 1 and 2, 2 and 3, and 1 and 3, as well as 30 average amplitude values. Subsequently, using Equation (6), we obtain 90 estimated AoA values for each data packet. We collected 2970 data packet samples at each training location and constructed 33 images of size 90$ \times $90 using the diverse data obtained. Each image represents the number of data packets in rows, and the columns represent the diversity data from the three antennas (the R, G, and B channels represent the average amplitude values, estimated AoA values, and phase values, respectively). Fig.~\ref{fig2} shows the RGB channel images of three different locations. Different distributions of RGB channel images can be observed at various locations, serving as distinctive fingerprints for indoor localization applications.

\section{The SECCI system}
\label{s3}

\subsection{Secci System Architecture}
\label{s3-1}
\begin{figure}[!ht]
	\centering
	\includegraphics[width=0.5\textwidth]{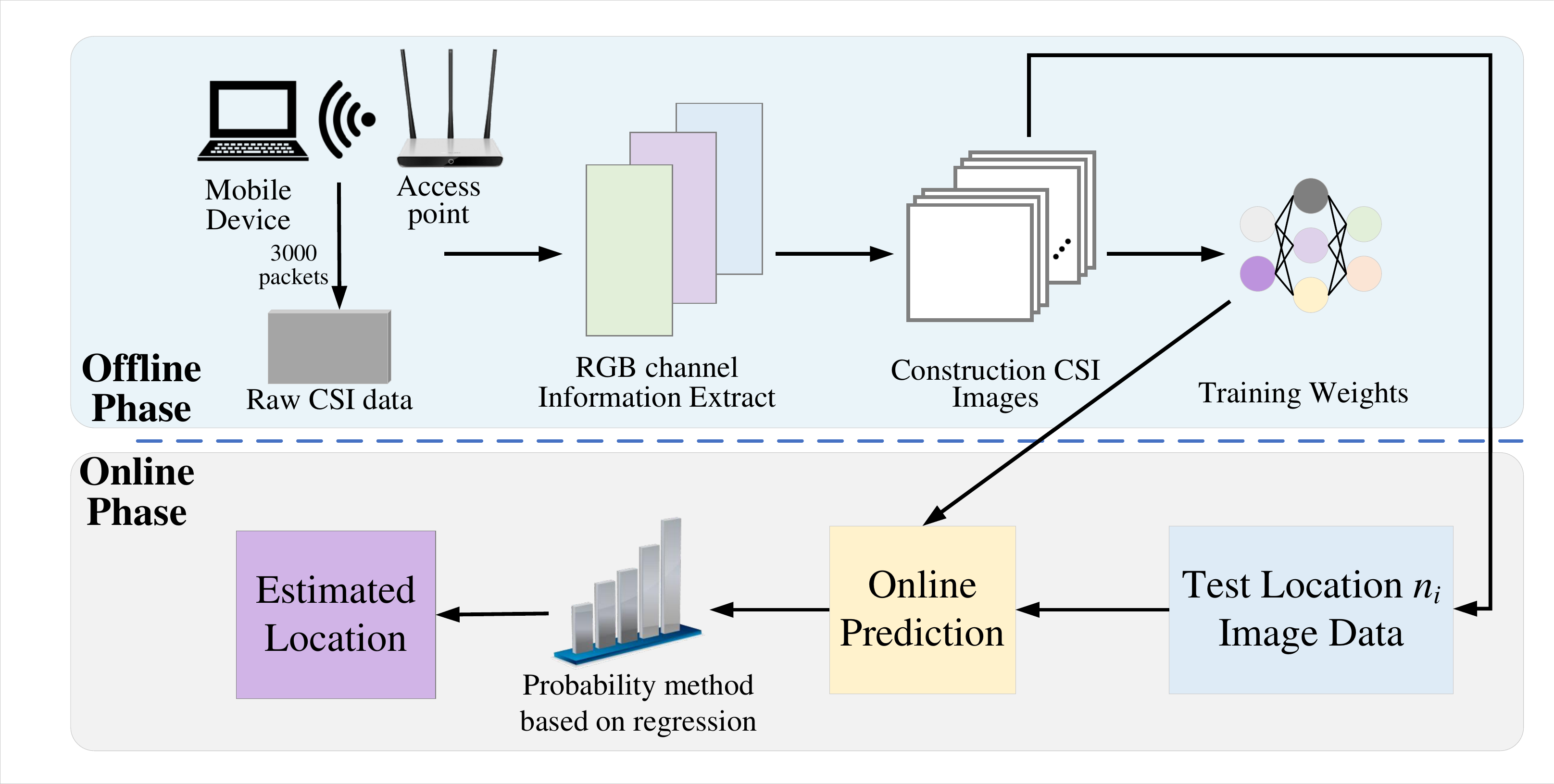}
	\caption{The Secci system architecture.}
	\label{fig33}
\end{figure}

Fig.~\ref{fig33} illustrates the architecture of the Secci system. The mobile device, represented by a laptop, communicates with the access point, represented by a desktop computer, using the Intel 5300 NIC in both devices. They are installed with the Linux 802.11n CSI Tool\cite{29} and collect CSI packets in Monitor mode. This mode offers greater stability compared to the AP mode, eliminating packet loss and allowing for adjustable packet transmission rates, effectively reducing the data collection time during the offline phase. Additionally, this mode allows for flexible configuration of physical layer parameters according to specific requirements. To ensure channel stability, data collection is conducted in the 5 GHz frequency band\cite{wang2017biloc}. At each location, data is collected for several seconds, with a minimum of 7000 data packets obtained. To ensure data accuracy, the initial 2 seconds of data are discarded, and the amplitude and phase information from the remaining 2970 data packets are used to construct CSI images.

One of the distinguishing features of the Secci system is its diversified data structure design. The Intel 5300 NIC has three antennas that exhibit distinct CSI characteristics, providing an opportunity to enrich the diversity of training and testing samples. Thus, we obtain three sets of data: (\romannumeral1) 90 average amplitudes from antenna pairs 1 and 2, 1 and 3, and 2 and 3; (\romannumeral2) 90 estimated angles of arrival (AOAs) from antenna pairs 1 and 2, 1 and 3, and 2 and 3; (\romannumeral3) 90 phase values from antenna 1,2, and 3. Secci utilizes CSI images primarily for the following reasons. Firstly, the three types of CSI data mentioned above exhibit significant stability in consecutive received packets at any given location. Secondly, in certain indoor environments, they can complement each other and enhance fingerprint stability. In scenarios where the WiFi signal encounters obstacles such as walls or furniture, the average amplitude experiences significant attenuation, while the estimated AoA values exhibit robustness with minimal impact. Phase data is also more resilient than amplitude data and contains many extractable features. Thirdly, CSI images can effectively utilize the different CSI characteristics of different antennas and the rich channel temporal and frequency characteristics present in all subcarriers of the received packets. We convert the extracted diversity features into images to enhance localization performance.

\subsection{Offline Training For Diverse Fingerprint Database}
\label{Design-of}

Typically, a CNN network consists of multiple convolutional layers, pooling layers, and fully connected layers. The convolution operation in CNN helps extract local features from images, capturing the local patterns in the data\cite{jiao2021overview}. This local perception makes CNN highly effective in processing visual and spatial data. The weight parameters in CNN are shared across different locations, reducing the parameter count and improving the training efficiency and generalization ability of the model. The hierarchical design of CNN allows it to learn more abstract and semantically rich features, enhancing the model's representational power, particularly in extracting features from diverse CSI data. Additionally, CNN often employs data augmentation techniques during training, such as random cropping, rotation, and scaling, to expand the training dataset. This helps increase the model's generalization ability and improve its robustness to input data variations. To enhance the feature representation capability of the network for CSI diversity data, we introduce the SE attention mechanism. The following sections will introduce the main components of Secci.

The convolution operation in CNN is used to extract translation-invariant features from input data. To compute the output feature map $Y_{p}$, the convolutional kernel $W_{p,1}, W_{p,2}, \cdots\,, W_{p,D}$ is convolved with input feature maps $X_{1}, X_{2}, \cdots\,, X_{D}$ individually. The convolution results are then summed together and added with a scalar bias $b$ to obtain the net input $Z^{p}$ of the convolutional layer. After passing through a nonlinear activation function, the output feature map $Y^{p}$ is obtained.

\begin{equation}
	Z^ p  = {W^p} \otimes {X} + {b^p} = \sum_{d=1}^D {W^{p,d}} \otimes {X^d} + {b^p} ,
	\label{eq27}
\end{equation}
\begin{equation}
	Y^p = \operatorname{\textit{f}}({Z^ p}).
	\label{eq28}
\end{equation}
where $W^{p}\in \mathbb{R}^{U\times V \times D}$ represents a three-dimensional convolutional kernel, and $\operatorname{\textit{f}}({\cdot})$ denotes the non-linear activation function. In this work, we apply the Rectified Linear Unit (ReLU) function \cite{31}. ReLU, also known as the Rectifier function, is a commonly used activation function in deep neural networks. It is essentially a ramp function defined as follows:
\begin{equation}
	\begin{aligned}
		y &= 
		\begin{cases}
			x, & x>0,\\
			0, & x \geqslant 0.
		\end{cases}\\
		&= max(0,x),
	\end{aligned}
	\label{eq3}
\end{equation}

It partially alleviates the problem of vanishing gradients in neural networks and accelerates the convergence speed of gradient descent. Batch Normalization (BN) is an effective layer-wise normalization method that can normalize any intermediate layer in a neural network. It is worth noting that layer-wise normalization not only improves optimization efficiency but also serves as a form of implicit regularization. It prevents neural networks from "overfitting" to specific samples, thereby enhancing the network's generalization ability.

The pooling layer serves to perform feature selection and reduce the number of features, thereby reducing the number of parameters and avoid overfitting. Although the convolutional layer can significantly reduce the number of connections in the network, the number of neurons in the feature map groups is not significantly reduced. To address this issue, a pooling layer can be added after the convolutional layer. Pooling refers to downsampling each region to obtain a single value that represents the summary of that region. In this paper, we use Max Pooling, which is defined as follows:
\begin{equation}
	y_{m,n}^d = \mathop{\max}_{i\in R_{m,n}^d} x_i,
	\label{eq28}
\end{equation}
where $x_i$ represents the activation value of each neuron in region $R_k^d$, $R_{m,n}^d$ denotes a specific region within each feature map, where $1\leq m \leq M$, $1\leq n \leq N $. Other methods such as summation or average pooling functions can also be used to accelerate training time. When training a deep neural network, dropout can be employed to prevent overfitting by randomly dropping a portion of neurons. dropout involves randomly selecting which neurons to drop during each training iteration. The simplest approach is to set a fixed probability $p$. For each neuron, it is determined to be kept or discarded independently with a probability of $p$.

The Fully Connected Layer is a common layer type in CNNs and plays a crucial role at the end of the network. It classifies and recognizes the features extracted from previous convolutional and pooling layers. The purpose of the Fully Connected Layer is to transform these abstract features into more specific categories or labels. The output of the Fully Connected Layer is a vector with a fixed dimension, where each element corresponds to a probability score for a category. The Softmax activation function is used to normalize these probabilities, ensuring that they sum up to 1, thereby representing the probabilities of the input belonging to each category. The discrepancy between the true position labels and the output data of SecciNet can be quantified using a loss function. By employing the Backpropagation (BP) algorithm and the AdamW optimizer \cite{32}, we aim to minimize the loss function and update the convolutional weights accordingly. The proposed SecciNet utilizes the cross-entropy loss function, which is defined as:
\begin{equation}
	\mathcal{L}{(\textbf{y},\,\operatorname{\textit{f}}\,(x;\theta))} = -\sum_{c=1}^{C} { y_c \, {\lg \, \operatorname{\textit{f}}_c(x;\theta)}} .
	\label{eq12}
\end{equation}
where $\textbf{y}$ represents the true distribution of labels, $\operatorname{\textit{f}}\,(x;\theta)$ denotes the predicted distribution by the model, $C$ is the dimension of the one-hot vector $y$, and $y_c$ corresponds to the true conditional probability of the coordinate being class $c \,(1\leq c\leq C)$.

\begin{figure*}[!ht]
	\centering
	\includegraphics[width=0.7\textwidth]{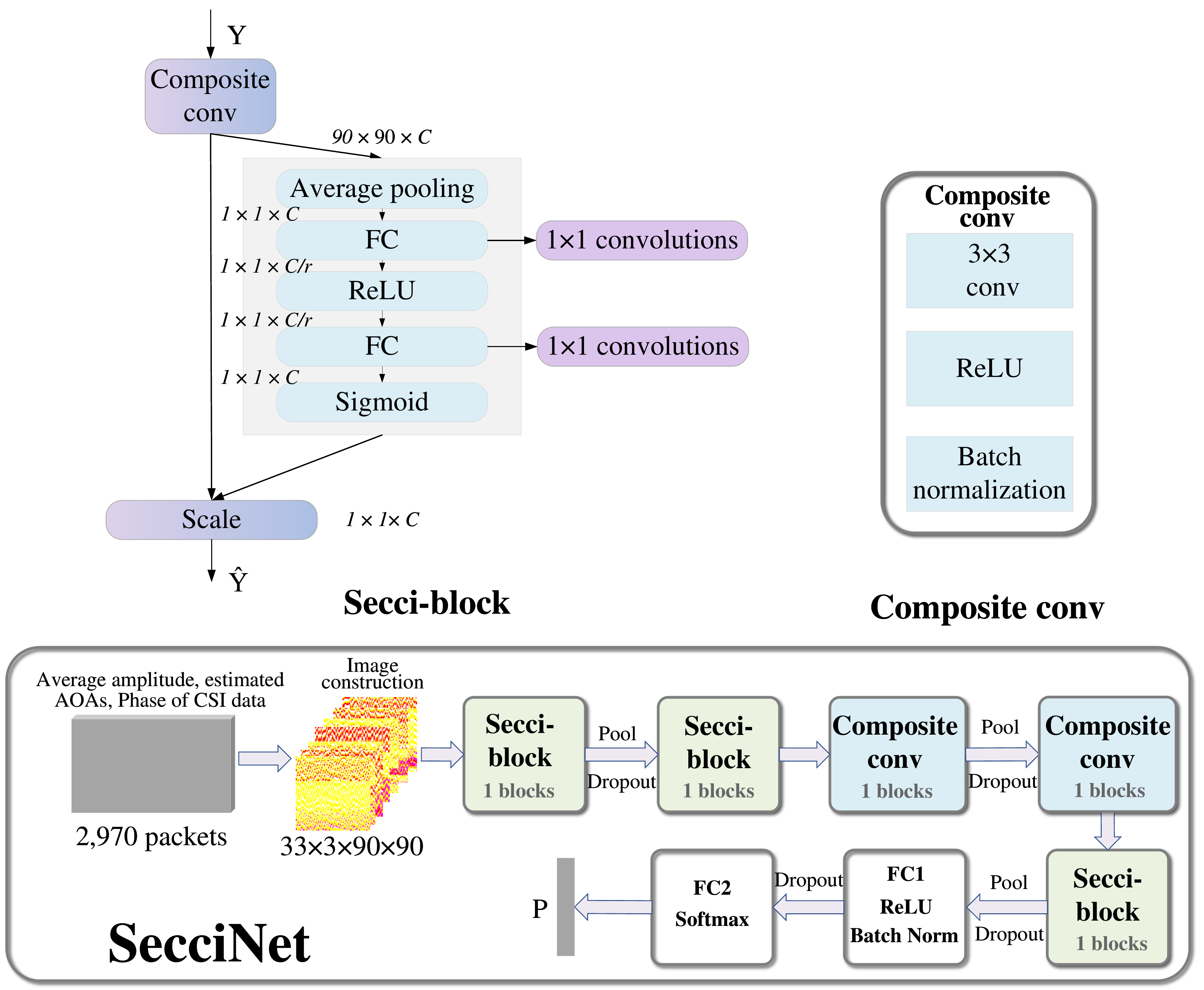}
	\caption{Network Architecture of SecciNet for CSI images training.}
	\label{fig4}
\end{figure*}

Squeeze-and-excitation networks (SE networks)\cite{33} are attention mechanisms used to enhance the representation capability of CNNs. The core idea is to introduce a component called the "squeeze-and-excitation" module into the feature maps of the CNN, which automatically selects and enhances the most discriminative feature channels. As shown in Fig.~\ref{fig4}, in SE networks, the input feature maps are spatially reduced through global average pooling, resulting in a global feature descriptor vector $\textbf{z}\in \mathbb{R}^{C}$. Specifically, the $c$-th element of $\textbf{z}$ is calculated by:
\begin{equation}
	z_c = \textbf{F}_{sq}(\textbf{u}_c) =\dfrac{1}{H \times W} \sum_{i=1}^{H}\sum_{j=1}^{W} {u_c(i,j)} .
	\label{eq13}
\end{equation}
where $\textbf{u}_c \!\in \!\mathbb{R}^{H\times W}$ represents a particular feature map $\textbf{U} = [\textbf{u}_1, \textbf{u}_2,...,\textbf{u}_C]$ after the Composite conv operation, $\textbf{F}_{sq}$ denotes the transformation of Global Information Embedding. $H$, $W$ represent the height and width of the feature map, respectively. Subsequently, a pair of densely connected fully connected layers are introduced to map the global feature descriptor vector to a weight vector. Here, we employ a convolutional layer with a kernel size of 1 and a simple gating mechanism, along with the use of the Sigmoid activation function, to achieve this mapping.
\vspace{1em}
\begin{equation}
	\textbf{s} = \textbf{F}_{ex}(\textbf{z},\textbf{W}) = \sigma(g(\textbf{z},\textbf{W})) = \sigma(\textbf{W}_2\delta(\textbf{W}_1\textbf{z})).
	\label{eq14}
\end{equation}
where $\delta$ denotes the ReLU function, $\textbf{W}_1\in \mathbb{R}^{{\frac{C}{r}}\times C}$ and $\textbf{W}_2\in \mathbb{R}^{C\times {\frac{C}{r}}}$. This weight vector corresponds to each channel of the input feature map and reflects the importance of each channel in capturing discriminative features. For each channel, the channel features are weighted by multiplying the weight vector with the input feature map.
\begin{equation}
	\widetilde{\textbf{x}}_c = \textbf{F}_{scale}(\textbf{u}_c,s_c) = s_c\textbf{u}_c .
	\label{eq15}
\end{equation}
where $\widetilde{\textbf{X}} = [\widetilde{\textbf{x}}_1, \widetilde{\textbf{x}}_2,..., \widetilde{\textbf{x}}_C]$ and $\textbf{F}_{scale}(\textbf{u}_c,s_c)$ refers to channelwise multiplication between the scalar $s_c$ and the feature map $\textbf{u}_c$. This approach helps the network to better capture the relevant information in the input location data, resulting in a significant improvement in the model's performance.

Fig.~\ref{fig4} illustrates the training process of CSI image based Secci.  Firstly, we construct images of size 90$ \times $90$ \times $3 using diverse features extracted from 2970 packets per sample point. Different sizes of three Secci-blocks and two Composite convolutions are employed to fully extract the latent features of each input image. To prevent overfitting during training, L2 regularizer, BN, and Dropout are introduced in the convolutional layers to optimize the training process. Data augmentation techniques such as RandomHorizontalFlip and RandomVerticalFlip are applied at the beginning of training. Finally, the feature maps pass through fully connected layers and are then processed by the softmax layer to obtain the output results. The weights are updated using the AdamW optimizer with the combination of position labels of training samples and the cross entropy loss function. AdamW optimizer is used to address the issue of L2 regularization failure in the Adam optimizer. The output results provide the probabilities of each coordinate category.

The offline training of Secci is outlined in Algorithms 1 through pseudocode. The input consists of training and testing images of size 90$ \times $90$ \times $3, corresponding labels, learning rate, number of iterations, and batch size. Initially, we perform preprocessing on the image data, including normalization and data augmentation, and randomly generate weights and biases (lines 1-3). The first and last layers represent the input layer and softmax output layer, respectively, while the intermediate layers consist of Secci-block and Composite conv. Samples are randomly selected, and the model is trained using the batch size for each epoch (lines 4-21). The convolutional layer outputs are linked to the fully connected layer, and the loss function is employed to calculate the discrepancies between the predicted and true labels for BP (lines 22-30). Finally, we obtain the trained weights (w) and biases (b), which are used to predict the test set of CSI images.

\begin{algorithm}[H]
	\caption{Secci training on CSI images.}\label{alg:alg1}
	\begin{algorithmic}
		\REQUIRE Training CSI images, Position labels, $Max\underline{~}epoch$, Regularization coefficient $\lambda$, Network architecture, and Learning rate $\alpha$.		
		\ENSURE Trained model with weights $w$ and biases $b$.
		\STATE Data augmentation; Randomly initialize weights $w$ and biases $b$; Set epoch = 0;
		\WHILE {epoch $<$ $Max\_epoch$}
		\STATE Randomly select a mini-batch from the CSI images and location labels;
		\FOR{each mini-batch}
		\STATE $//$ Forward propagation;
		\STATE Set input data as the CSI images;
		\FOR{each layer $l$ from $2$ to $L-1$}
		\IF{the current layer $l$ is a convolution layer in Secci-block} 
		\STATE Calculate the feature map according to $(1-8)$;
		\ELSE
		\STATE $//$ The current layer is a Composite conv;
		\STATE Calculate the feature map according to $(1-5)$;
		\ENDIF 
		\ENDFOR
		\STATE $//$ The last layer is a pair fully-connected layer;
		\STATE Flatten the feature maps;
		\STATE  $v = Dence(\theta^{L-1})$; $o = \sigma(w^L \times v + b^{L})$;
		\STATE  Apply the activation function;
		\STATE $//$ Loss function;
		\STATE  Calculate the loss between the output and the location labels according to (5);
		\STATE  $//$ Backpropagation;
		\STATE  Update the weights and biases using the BP algorithm;
		\STATE  $\dfrac{\partial {\mathcal{L}(\textbf{y}^{(n)}, \hat{\textbf{y}}^{(n)})} }{\partial \mathcal{\textbf{W}}} = \delta^{(l)}{(\textbf{\textit{a}}^{(l-1)})^\mathsf{T} } $; $\dfrac{\partial {\mathcal{L}(\textbf{y}^{(n)}, \hat{\textbf{y}}^{(n)})} }{\partial \textit{\textbf{b}}^{(l)}} = \delta^{(l)} $;
		\STATE $//$ Update Parameters
		\STATE $\textbf{W}^{(l)} \leftarrow \textbf{W}^{(l)} - \alpha(\delta^{(l)} (\textbf{\textit{a}}^{(l-1)})^\mathsf{T} + \lambda \textbf{W}^{(l)} ) $; $\textbf{b}^{(l)} \leftarrow \textbf{b}^{(l)} - \alpha \delta^{(l)} $;
		\ENDFOR
		\STATE epoch = epoch + 1;
		\STATE Store network parameters with the highest accuracy for Validation;
		\ENDWHILE
	\end{algorithmic}
	\label{alg1}
\end{algorithm}

\subsection{Online Prediction Algorithm}
\label{Online Algorithm}

In this phase, we utilize the trained SecciNet and a probabilistic regression method to estimate the coordinates of the mobile device's location based on newly received CSI images. Assuming $N$ denots the total number of images obtained from all test positions, and $L$ represents the number of training locations. The network output for location $i$ using image $j$ is denoted as $p_{ij}$. To obtain the network output for the $L$ training locations using the $N$ images, we construct a matrix $\textbf{\textit{P}}$ as follows:
\vspace{1em}
\begin{equation}
	\textbf{\textit{P}}={
		\left[ \begin{array}{ccccc}
			p_{11} & p_{12} & p_{13} & \cdots & p_{1N}\\
			p_{21} & p_{22} &  p_{23} &  \cdots & p_{2N}\\
			\vdots & \vdots & \vdots & \ddots & \vdots\\
			p_{L1} & p_{L2} & p_{L3} & \cdots & p_{LN}
		\end{array} 
		\right ]},
	\label{eq9}
\end{equation}

For matrix $\textbf{\textit{P}}$, a proposed greedy method is utilized to select $H$ candidate locations for each data and compute the regression of these locations as the estimated location of the mobile device.  Firstly, $H$ largest output location indexes are chosen from each column of matrix $\textbf{\textit{P}}$, resulting in a new matrix R with size $N \times H$ as

\begin{equation}
	\textit{\textbf{R}}={
		\left[ \begin{array}{cccccc}
			r_{11} & r_{12} & \cdots & r_{ij} & \cdots & r_{1N}\\
			r_{21} & r_{22} & \cdots & r_{2j} &  \cdots & r_{2N}\\
			\vdots & \vdots & \vdots & \vdots & \ddots & \vdots\\
			r_{H1} & r_{H2} & \cdots & r_{Hj} &  \cdots & r_{HN}
		\end{array} 
		\right ]}.
	\label{eq10}
\end{equation}
where $r_{ij} \in \left\{1, 2, ..., L\right\}$is the location index of the $i$-th largest output for image $j$. Through the computation of frequencies for the same location index across matrix $\textbf{\textit{R}}$, we obtain $K$ sets of the highest location indexes. Moreover, the weight of the location index $i$ corresponds to the softmax output of the network and is denoted as $pi$. Finally, the estimated coordinates are obtained by taking the average of the $K$ selected location coordinates:

\begin{equation}
	\hat{\textbf{\textit{L}}} = \dfrac{1}{K} \sum_{j=1}^{K}\sum_{i=1}^{H} {l_i \times r_{ij}} ,\; K \:\textless \:N .
	\label{eq11}
\end{equation}
where $l_i$ represents the $i$-th indexed training location. In our experiments, we let $H=5$ to achieve stable and improved localization performance.

The pseudocode for online prediction of Secci is presented in Algorithms 2. The input consists of test images, corresponding labels, network architecture and parameters, $H_{num}$, and the number of training samples. First, we calculate the top $H_{num}$ position indexes with the highest probabilities for each data point (lines 1-2). Then, we compute the regression position coordinates for each sample. Finally, we calculate the frequencies of each training point data and obtain the expected regression of all sampled positions (lines 4-9).
\begin{algorithm}[H]
	\caption{Secci system location prediction.}\label{alg:alg2}
	\begin{algorithmic}
		\REQUIRE Test samples, location labels, network architecture, system parameter settings, $H_{num}$, $class_{num}$.		
		\ENSURE Estimated position coordinates.
		\STATE set $i,j = 0$;
		\FOR{$j$\, : \,$N$}
		\STATE Calculate the position indexes of the top H maximum probabilities for each data point $r_{*j}$;
		\FOR{$i$\, : \,$H_{num}$}
		\STATE $//$ Compute the regression-based position coordinates; 
		\STATE $L^{\prime} = \sum_{i=1}^{H} p_{ij} \times r_{ij}$;
		\ENDFOR
		\ENDFOR
		
		\FOR{$i$\, : \,$class_{num}$}
		\STATE Compute the expected regression for all sampled locations according to (11). 
		\ENDFOR
	\end{algorithmic}
	\label{alg2}
\end{algorithm}

\section{EXPERIMENTS AND PERFORMANCE ANALYSIS}
\label{s4}
In this section, we conducted detailed evaluation experiments from several aspects. We compared the proposed Secci algorithm with other indoor localization algorithms based on CSI images, namely ILCL\cite{34}, CNN5\cite{34}, BLS\cite{35}, CiFi\cite{36}.

\subsection{Experimental configuration}
\label{Simulation-results}

We utilized a Lenovo desktop computer as the access point and a Lenovo laptop as the mobile device to collect data packets containing CSI. Both devices were equipped with Intel 5300 NICs and the Linux 802.11n CSI Tool, and data packet collection was performed on the 64-bit Ubuntu desktop 12.04 LTS operating system. The physical layer was configured as 0X4101, with a guard interval of 0.8 $\upmu$s, channel bandwidth of 20 MHz, and an OFDM system with QPSK modulation and 1/2 coding rate. We conducted the data packet collection in a more stable Monitor mode for both the access point and the mobile device, selecting channel 116. At the access point, the transmission rates were set to 1000 packets/s in injection mode, supported by the LORCON version 1 for data packet injection. The distance between two adjacent antennas was $d = 2.68$ cm, which precisely corresponds to half the wavelength of the WiFi signal in the 5.58 GHz frequency band. With the aforementioned configuration, we received data packets from the mobile device's receiving NIC and extracted the CSI data.

To validate the system performance and compare different approaches, our experimental scenarios consisted of the following two indoor environments.

\begin{figure}[!ht]
	\setlength{\abovecaptionskip}{-0.1cm}   
	\centering
	\includegraphics[width=0.4\textwidth]{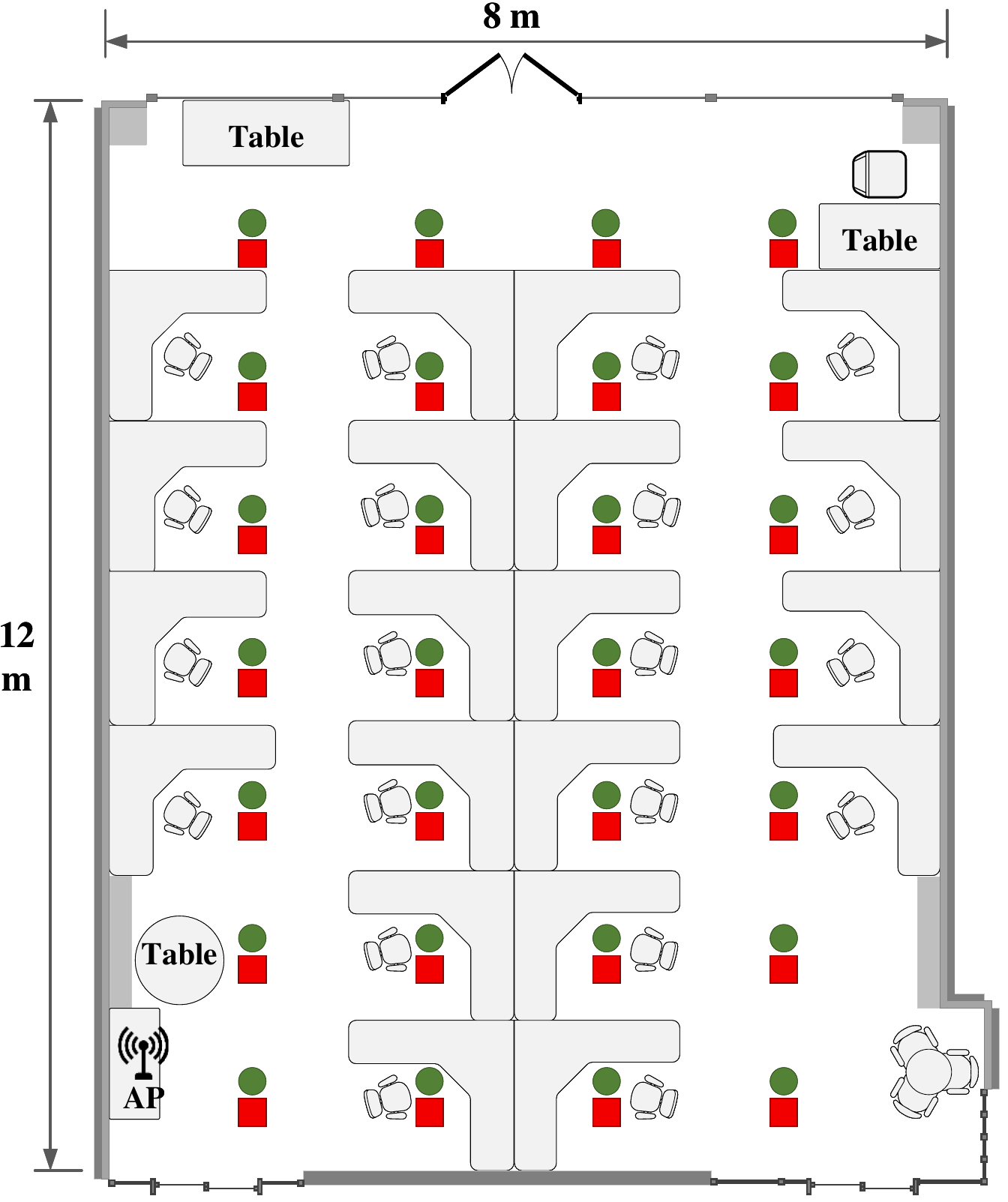}
	\caption{Training locations are indicated by green dots, while testing locations are represented by red squares, in the lab layout. }
	\label{fig5}
\end{figure}

\begin{figure}[!ht]
	\setlength{\abovecaptionskip}{-0.1cm}   
	\centering
	\includegraphics[width=0.5\textwidth]{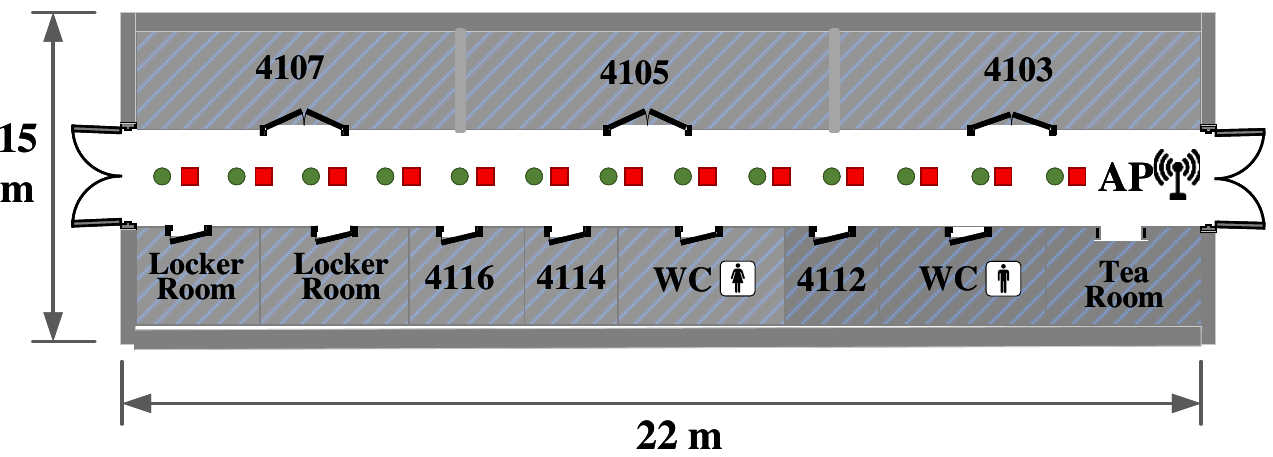}
	\caption{Training locations are denoted by red squares, and testing locations are represented by green dots, in the corridor layout.}
	\label{fig6}
\end{figure}

\emph{1) The Lab:} This is a 12$ \times $8 $m^{2}$ lab located in the China Wireless Valley, Nanjing City, within the School of Information Science and Engineering at Southeast University. The indoor environment is complex, with numerous desks, chairs, desktop computers, and electronic devices. Particularly, while collecting and creating the dataset, there are people working, and there is also frequent movement within the space. As a result, the environment contains a significant number of NLOS paths, making it a typical NLOS experimental scenario, as illustrated in Fig.~\ref{fig5}. The scenario comprises 56 sampling positions, with half of them designated as training samples (marked with green dots), and the remaining positions as test locations (marked with red squares). The distance between adjacent training positions is 1.5 m. The transmitter is placed on the table in the corner, while the receiver remains at the same height to obtain 3000 data packets from both the training and test positions. Each position is associated with 33 CSI image samples.

\emph{2) The Corridor:} This is a 12$ \times $24 $m^{2}$ long corridor located in China Wireless Valley. The indoor environment of this corridor is spacious, and there is minimal movement during the data collection process. Therefore, this environment serves as a typical LOS experimental scenario, as depicted in Fig.~\ref{fig6}. The scenario includes 26 sampling positions, with half of them designated as training samples (marked with green dots), and the remaining positions as test locations (marked with red squares). The distance between adjacent training positions is 1.6 m. The transmitter is placed on the floor at the end of the corridor, while the receiver remains at the same height to collect data from both the training and test positions for a few seconds. We collect 3000 data packets during the intermediate collection time to ensure the validity of the CSI data. Each position is associated with 33 CSI image samples.

%

\begin{table}[]
	\centering
	\caption{LOCALIZATION ACCURACY AND TESTING TIME(lab, $H = 5$).}
	\label{tab1tab2}
	\begin{tabular}{ccccccc}
		\toprule
		\multirow{2}{*}{\textbf{Algorithm}}    & \multicolumn{3}{c}{\textbf{Lab}} & \multicolumn{3}{c}{\textbf{Corridor}}            \\ \cline{2-7} 
		
		& \makecell[c]{\textbf{Mean error}\\\textbf{(m)}}    & \makecell[c]{\textbf{Std}\\\textbf{(m)}}& \makecell[c]{\textbf{Mean execution time}\\\textbf{(s)}} & \makecell[c]{\textbf{Mean error}\\\textbf{(m)}}& \makecell[c]{\textbf{Std}\\\textbf{(m)}}& \makecell[c]{\textbf{Mean execution time}\\\textbf{(s)}} \\ \midrule
		
		\multicolumn{1}{c}{\textbf{Secci}} & \textbf{2.775572 }   & 1.579896  & 0.056899 &  \textbf{2.308345 }& \textbf{2.450992} & 0.048790                  \\ 
		\multicolumn{1}{c}{ILCL} & 3.195439  &1.658585       & 0.059357            &   3.885138    &  2.974207  & 0.049247  \\ 
		\multicolumn{1}{c}{CNN5} & 3.214566  & 2.197467          & 0.051823        &     2.629091   & 3.102518  &  0.046205  \\ 
		\multicolumn{1}{c}{BLS} & 3.445957   & \textbf{1.430732}    & \textbf{0.007534}&   5.014658  & 2.584734  &  \textbf{ 0.003042}             \\ 
		\multicolumn{1}{c}{CiFi} & 3.123977  & 1.687454             & 0.008078         &   2.72979    &  2.566218  &   0.006993   \\ \bottomrule
	\end{tabular}
\end{table}


We implemented the Secci system on python, sklearn, and Baidu's Paddle framework, and Matlab was used for CSI data extraction and construction of CSI images. The Secci system runs on Baidu's AIStudio, and its configuration is 2 Intel(R) Xeon(R) Gold 6271C CPUs, 16GB RAM, 1 Tesla V100-SXM2-32GB GPU, 16GB video memory, and 100GB disk.

The mean positioning error is the main indicator of positioning performance, it is given by
\begin{equation}
	{d_i} = \frac{1}{{N}}\sum_{i=0}^N\sqrt{\left(\widehat{x}_{i} - {x_i}\right)^2 + \left(\widehat{y}_{i} - {y_i}\right)^2}.
	\label{eq25}
\end{equation}
Where $(\widehat{x}_{i},\widehat{y}_{i})$ is the estimated position coordinates, $(x_i,y_i)$ is the actual position coordinates.

\vspace{-0.8cm}
\subsection{Positioning accuracy and comparison}
\label{Simulation-results}
Table~\ref{tab1tab2} display the average errors, STD, and mean execution times for five schemes in two experimental scenarios. The proposed Secci scheme achieves an average error of 2.775572 m with a standard deviation of 1.579896 m in the lab environment. In the corridor environment, where NLOS transmission is reduced and the data is cleaner, Secci achieves an average error of 2.308345 m with a standard deviation of 2.450992 m, demonstrating improved accuracy. Secci outperforms the other four schemes in both scenarios. This is primarily attributed to Secci's utilization of diverse and complementary multi-source data, which enhances stability and robustness in complex environments. Additionally, the network architecture design enhances the network's perception of discriminative features, allowing for better capture of relevant information from input data. The execution time is largely dependent on the hardware environment and DL framework employed. Due to the online phase's image construction and the unique mechanisms of the Broad Learning System, Secci exhibits relatively longer average execution time compared to the other schemes. In the lab and corridor scenarios, Secci's average execution times are 0.056899 s and 0.048790 s, respectively, which are sufficient for achieving real-time localization.

\begin{figure*}[!ht]
	\centering
	\vspace{-0.2cm}  
	\setlength{\belowcaptionskip}{-6pt} 
	\setlength{\belowdisplayskip}{1pt} 
	\subfloat[\label{fig:arm1} {\small Lab}]{\includegraphics[width=0.4\linewidth]{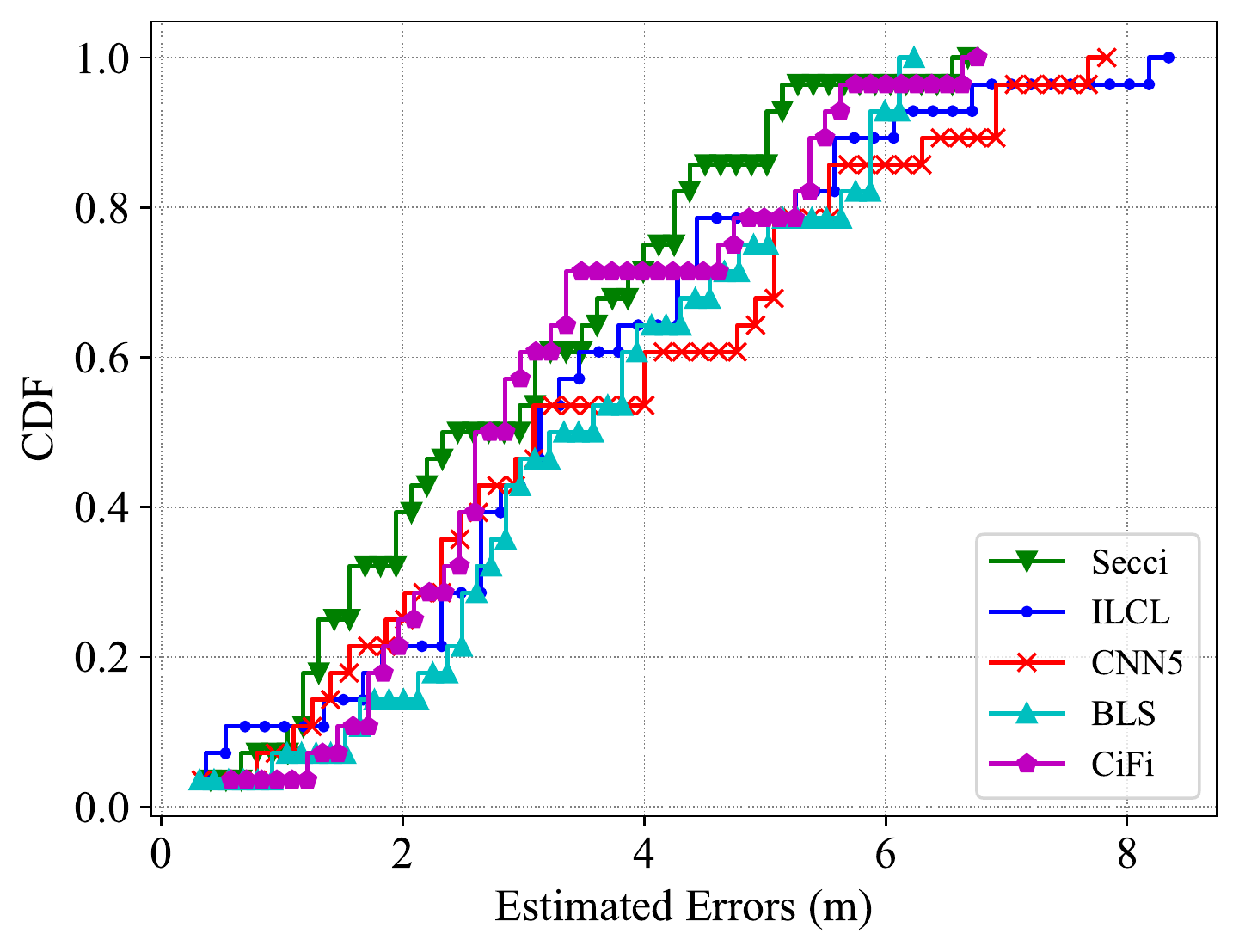}} \hspace*{0.1px} 
	\subfloat[\label{fig:arm1}{\small Corridor}]{\includegraphics[width=0.4\linewidth]{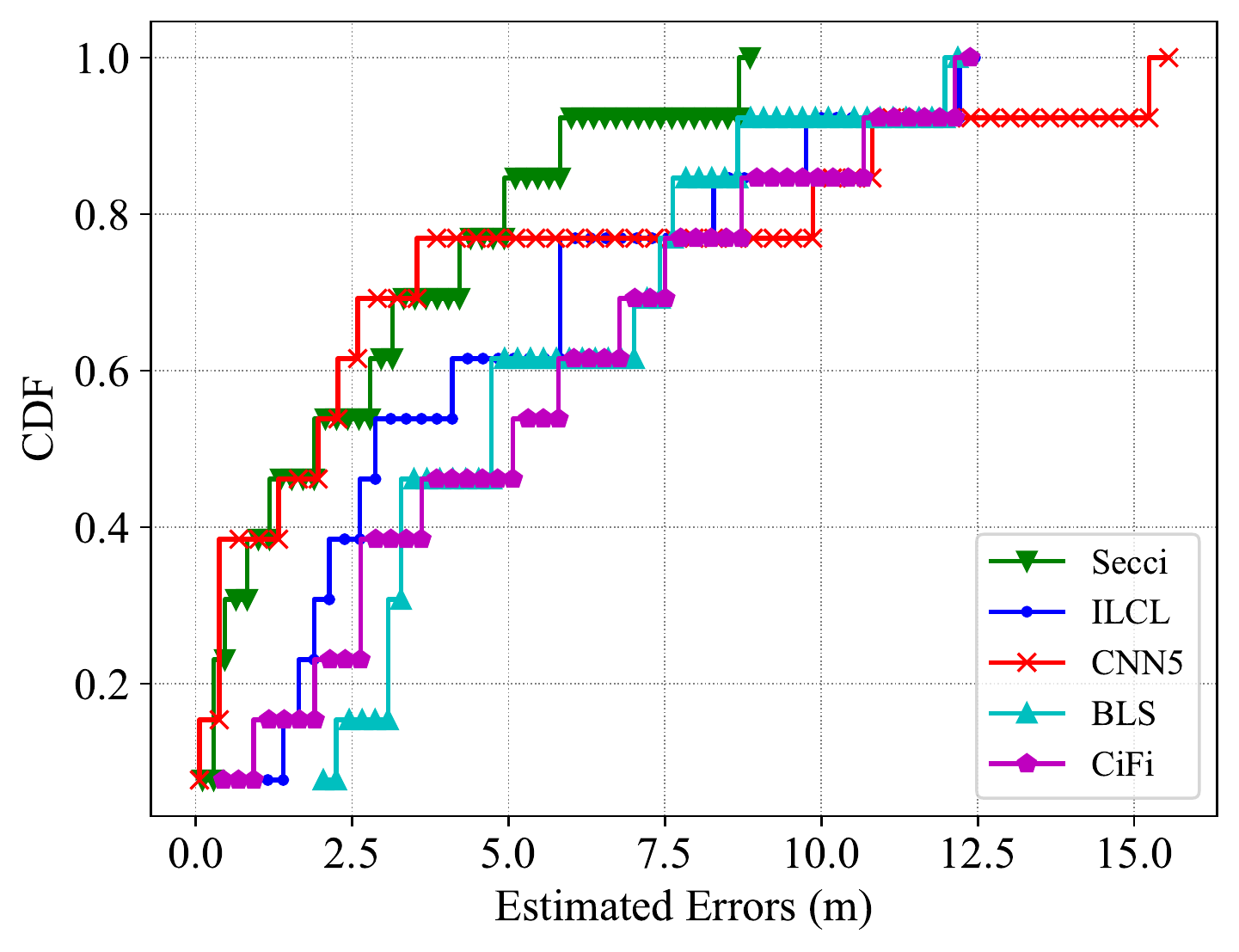}} 
	\caption{CDF of estimated errors for the lab and corridor experiment.}
	\label{fig7fig8}
\end{figure*}

As shown in Fig.~\ref{fig7fig8}(a), we evaluate the localization errors using the cumulative distribution function (CDF). In the lab environment, there is rich multipath information, such as people walking around. However, Secci can achieve high-precision localization by leveraging unique multipath features. Secci has 40$\%$ of the test locations with an average error within 2m, while the other schemes have 21$\%$, 25$\%$, 15$\%$, and 25$\%$, respectively. Furthermore, approximately 60$\%$ of the test locations have errors less than 3m, which is comparable to CiFi but outperforms the other three schemes. Therefore, in this experiment, Secci demonstrates the best performance. This is because the constructed diversified data is more robust to indoor multipath environments, and the designed network can better learn discriminative features, resulting in improved localization performance.

Fig.~\ref{fig7fig8}(b) shows the CDF of localization errors for each scheme in the corridor scenario. In this scenario, there is less signal shielding, and the data is cleaner, leading to slightly improved accuracy. Approximately 39$\%$ of the Secci test locations have errors less than 1m, while the rates for ILCL, CNN5, BLS, and CiFi are 5$\%$, 38$\%$, 0$\%$, and 15$\%$, respectively. Furthermore, the maximum error for Secci is 8.75m, while the other schemes have maximum errors exceeding 12m. This validates that the Secci system is more robust than the other four schemes, and Secci achieves higher accuracy with only one access point.

\begin{figure}[!ht]
	\centering
	\vspace{-0.2cm}  
	\setlength{\belowcaptionskip}{-6pt} 
	\setlength{\belowdisplayskip}{1pt} 
	\subfloat[\label{fig:arm1} {\small Training accuracy}]{\includegraphics[width=0.38\linewidth]{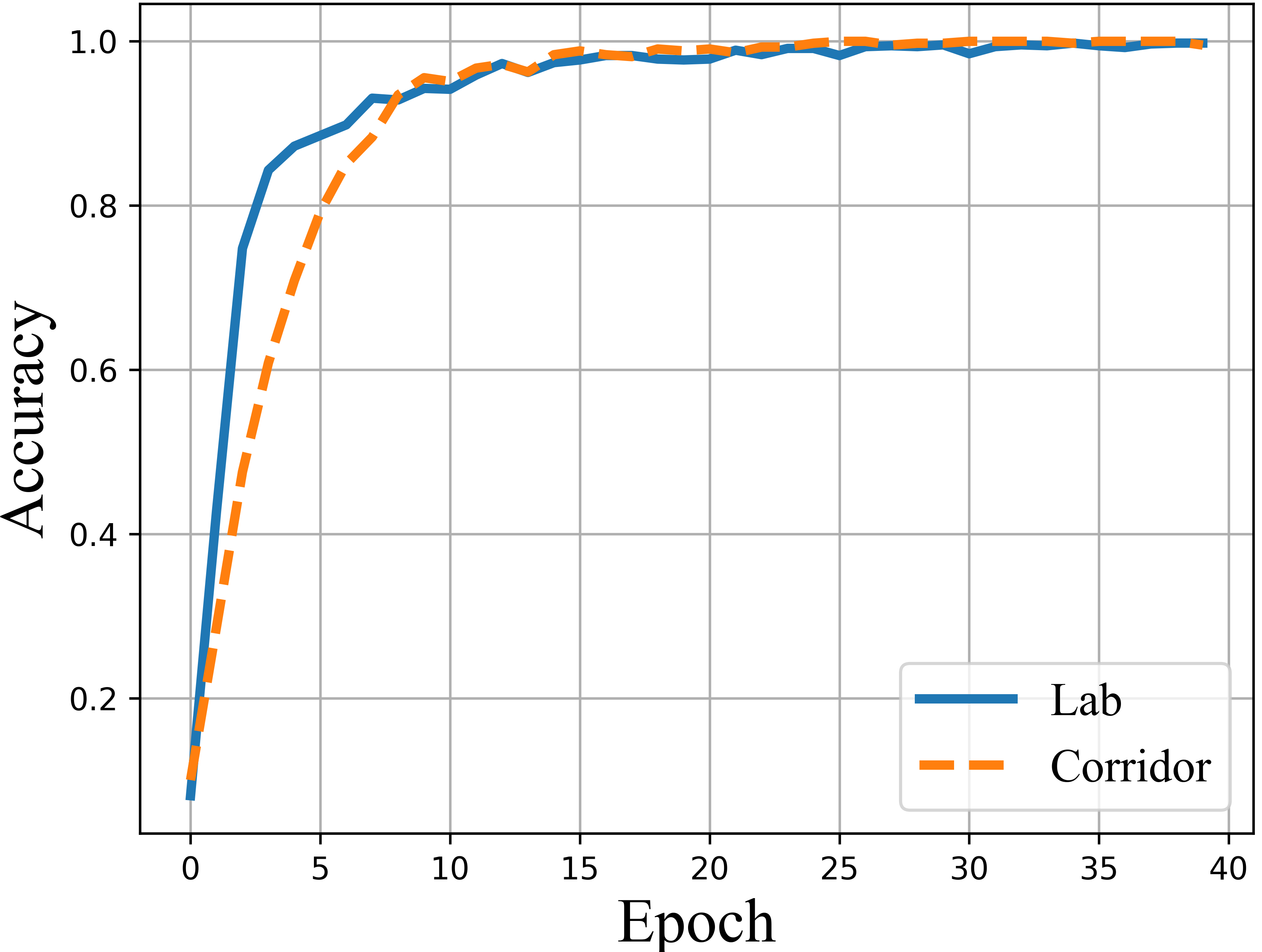}} \hspace*{0.1px} 
	\subfloat[\label{fig:arm1}{\small Training loss}]{\includegraphics[width=0.38\linewidth]{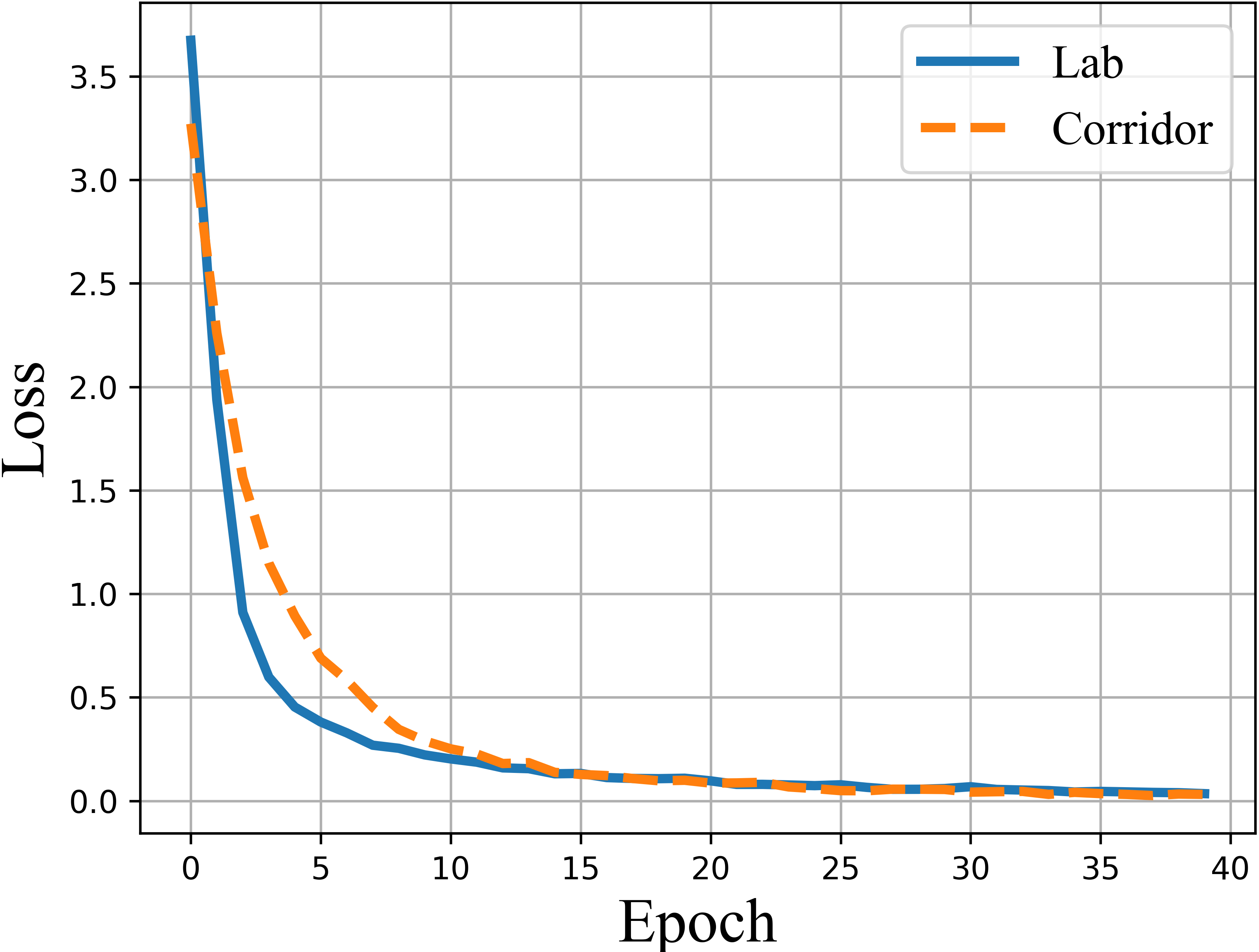}} 
	\caption{Training Accuracy and loss of Secci for the experimental scenarios.}
	\label{accloss}
\end{figure}
Fig.~\ref{accloss} illustrates the variation of training accuracy and loss over epochs in the lab and corridor environments. We set the learning rate to 3e-4 and the number of epochs to 40 to ensure training performance and avoid overfitting. During training, we store the model parameters from the epoch with the highest validation accuracy. Specifically, there are 33 CSI images per sampling point, which accelerates the convergence of the model and reduces the number of training iterations, thus saving time costs. As shown in Fig.~\ref{accloss}(b), the training loss decreases more rapidly in the lab scenario compared to the corridor scenario, but they exhibit similar convergence behavior, starting to converge around 30 epochs and stabilizing around a loss value of approximately 0.07.

\subsection{Effect of Different System Parameters}
\label{Simulation-results}
\subsubsection{Effect of Number of Training Images}
\label{s4-3-1}

In this subsection, we investigated the effect of different numbers of training images. To this end, we constructed five training and testing sets of varying sizes in both the lab and corridor environments, including each training location. To ensure fairness, the testing data were collected near the training points, and CSI images were generated using 90 average amplitude values, 90 estimated AOA values, and 90 phase values obtained from three antennas. Furthermore, other system parameters remained consistent, such as the number of reference positions ($H = 5$), batch size (Batchsize $ = 50$), and learning rate ($\alpha = 3e-4$).

\begin{figure}[!ht]
	\setlength{\abovecaptionskip}{-0.1cm}   
	\centering
	\includegraphics[width=0.42\textwidth]{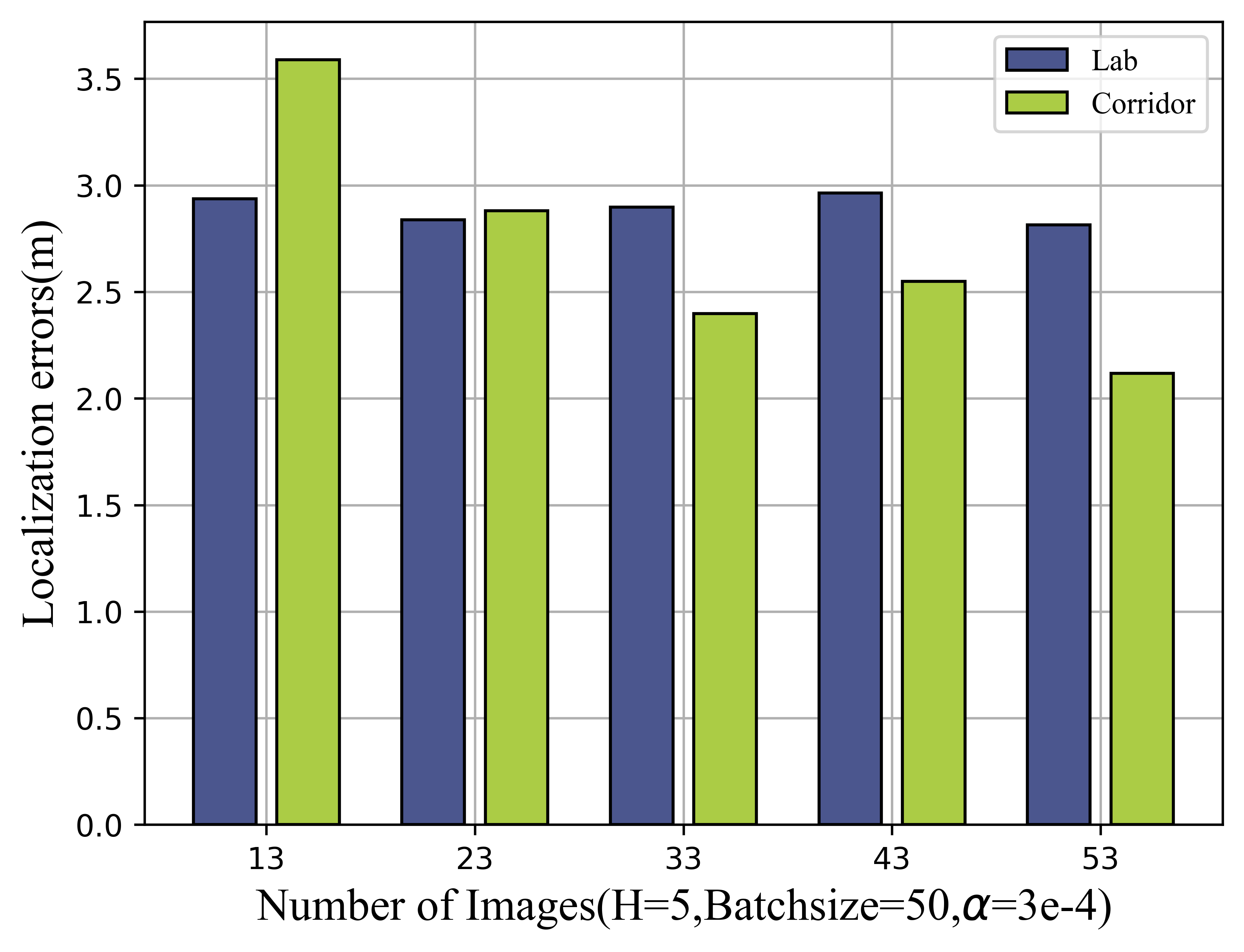}
	\caption{Mean localization errors were evaluated in the lab and corridor environments using varying numbers of training images.}
	\label{fig9}
\end{figure}

Fig.~\ref{fig9} demonstrates that increasing the number of images per training location leads to a decrease in average localization error in both the corridor and lab environments. In the corridor environment, the average localization error shows a consistent decrease trend. In the lab environment, the average localization error also decreases, albeit with minor fluctuations. The range of localization errors ranges from 2.814 m to 2.964 m. This indicates that Secci can learn robust fingerprint features through small-sample learning. The accuracy achieved in both scenarios is noteworthy for most location-based services. Our proposed Secci system can pursue better accuracy by increasing the dataset size. However, acceptable accuracy can still be attained with a small dataset when storage costs or data collection time are limited.

\subsubsection{Effect of Batchsize}
\label{s4-3-2}
In this experiment, we considered the impact of different batch sizes on the average localization error in two scenarios. To ensure fairness, we set the other parameters as follows: the number of reference positions ($H = 5$), the number of training images (num\_image $ = 33$), and the learning rate ($\alpha = 20$).
\begin{figure}[!ht]
	\setlength{\abovecaptionskip}{-0.1cm}   
	\centering
	\includegraphics[width=0.42\textwidth]{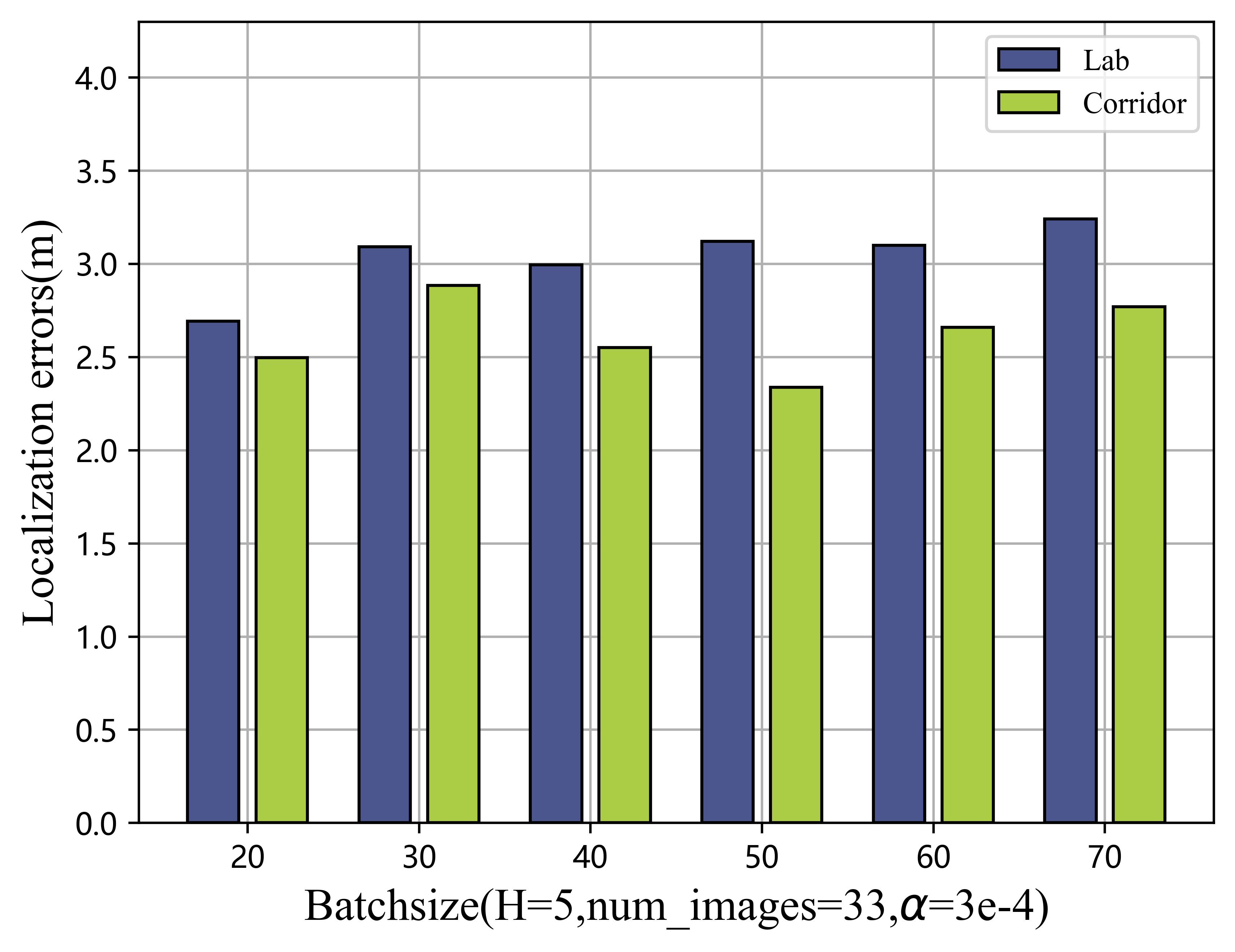}
	
	\caption{Mean localization errors were assessed in the lab and corridor environments with different batch sizes.}
	\label{fig10}
\end{figure}

Fig.~\ref{fig10} shows that as the batch size increases, the average error in the lab scenario initially rises but subsequently exhibits minor variations. The minimum error is achieved at a batch size of 20, with the maximum and minimum average errors being 3.241m and 2.691m, respectively. In contrast, the average localization error in the corridor scenario initially decreases and then increases, reaching a minimum at a batch size of 50, with the maximum and minimum average errors being 2.884m and 2.337m, respectively. The batch size corresponds to the "forest" and "trees" analogy. In the lab scenario, where multipath characteristics are prominent, excellent performance can be achieved even with a small batch size. On the other hand, the corridor scenario primarily involves LOS transmission, requiring a larger batch size to achieve higher accuracy. For the sake of comparison and discussion, we set the batch size to 50 in both scenarios.
\subsubsection{Effect of Learning Rate $\alpha$}
\label{s4-3-3}

Next, we investigated the impact of different learning rates on localization precision. In the experiment, we kept the other parameters as follows to ensure fairness: the number of reference positions ($H = 5$), the number of training images (num\_image $ = 33$), and the batch size (Batchsize $ = 20$).

\begin{figure}[!ht]
	\setlength{\abovecaptionskip}{-0.1cm}   
	\centering
	\includegraphics[width=0.42\textwidth]{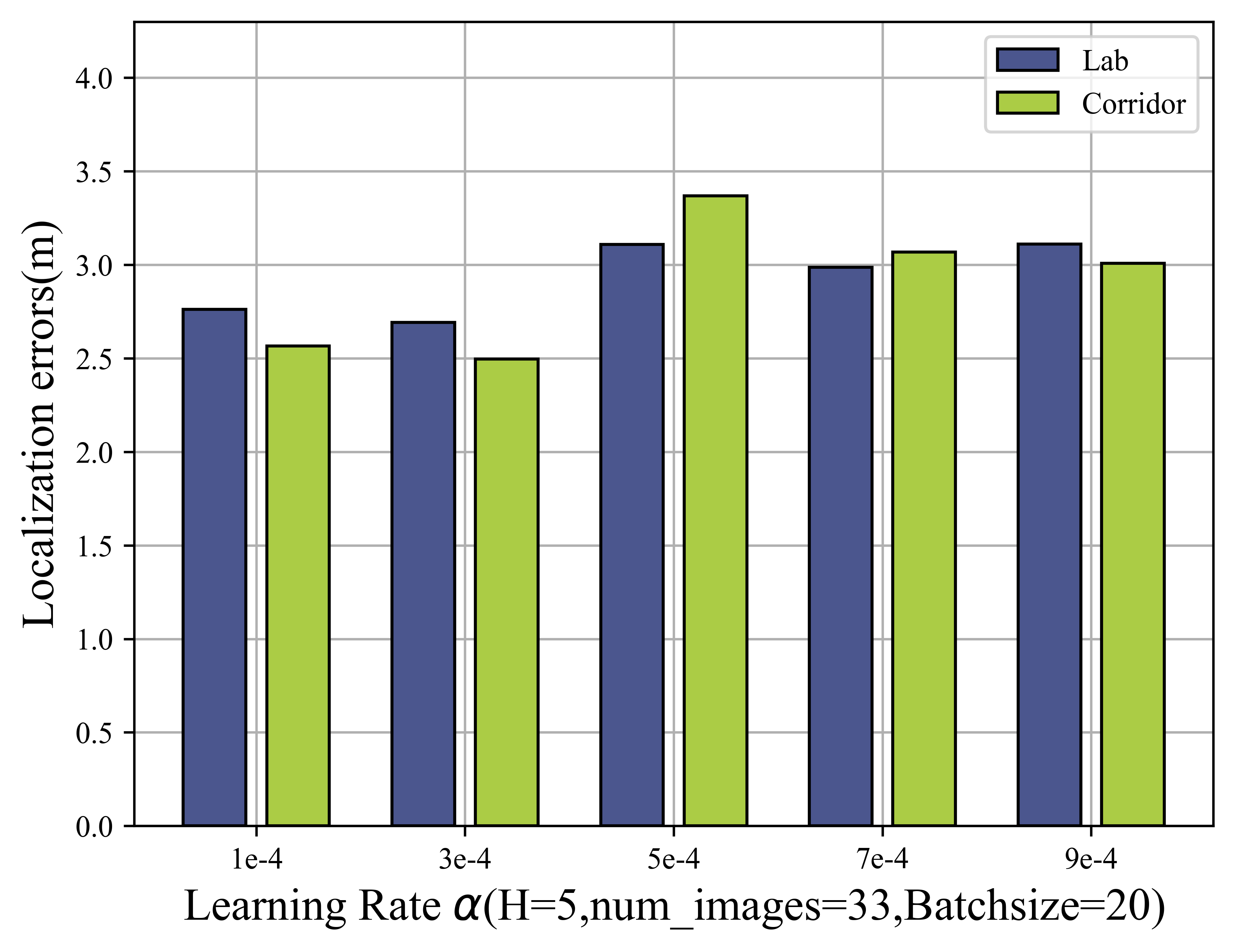}
	\caption{Mean localization errors were analyzed in the lab and corridor environments across different learning rates.}
	\label{fig11}
\end{figure}

Fig.~\ref{fig11} illustrates the average localization errors for different learning rates in the two scenarios. As the learning rate increases, the localization error initially decreases and then increases in both scenarios. The minimum error is achieved at a learning rate of 3e-4, while the maximum error is reached at a learning rate of 5e-4, with the corridor scenario exhibiting higher error than the lab scenario. This indicates that a relatively high learning rate prevents the network from achieving optimal convergence performance as the BP algorithm repeatedly hops back and forth over the valley. Therefore, to pursue the lowest localization error, the learning rate for both scenarios is set to 3e-4.
\subsubsection{Effect of Reference Locations $H$}
\label{s4-3-4}

In this section, we investigated the influence of H on the mean localization error in the Secci system. We employed a greedy method to select $H$ reference positions and used their expected regression as the estimated location. To ensure fairness, we set the other parameters as follows: the learning rate ($H = 5$), the number of training images (num\_image $ = 33$), and the batch size (Batchsize $ = 20$).

\begin{figure}[!ht]
	\setlength{\abovecaptionskip}{-0.1cm}   
	\centering
	\includegraphics[width=0.42\textwidth]{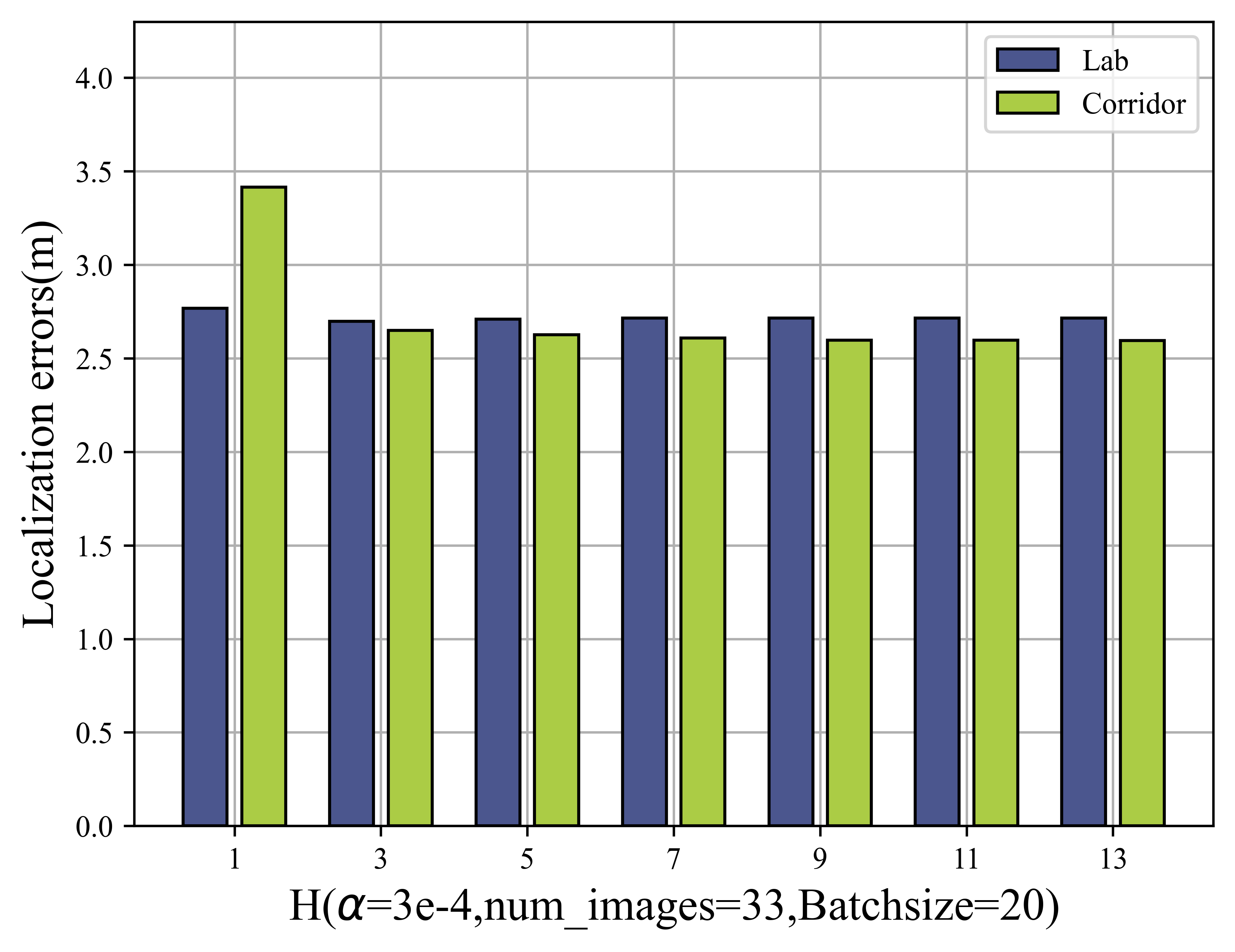}
	\caption{Mean localization errors were measured in the lab and corridor environments for various H values.}
	\label{fig12}
\end{figure}

From Fig.~\ref{fig12}, it can be observed that in the lab scenario, the mean localization error slightly decreases as the value of $H$ increases. However, once $H$ exceeds 5, the mean localization error remains nearly the same. In the corridor scenario, the mean localization error decreases significantly with the increase in $H$, and similarly, once $H$ exceeds 5, the mean localization error becomes nearly constant. This indicates that Secci is robust to the value of $H$. Therefore, to save costs and improve efficiency, we choose to utilize the top five outputs for location estimation.

\begin{figure}[!ht]
	\setlength{\abovecaptionskip}{-0.1cm}   
	\centering
	\includegraphics[width=0.42\textwidth]{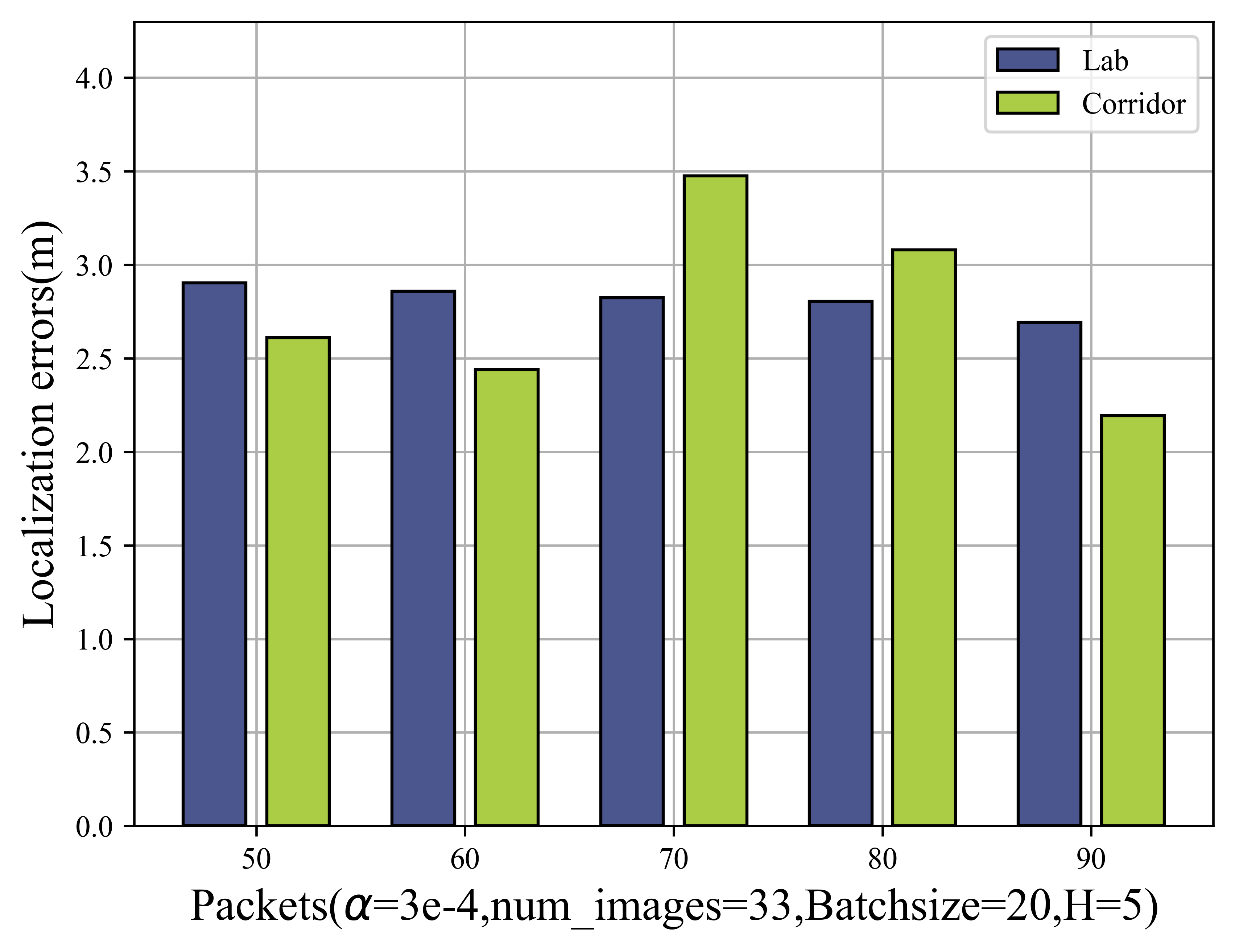}
	\caption{Mean localization errors were determined in the lab and corridor environments for different numbers of packets in training and testing images.}
	\label{fig13}
\end{figure}
\subsubsection{Effect of Number of Training Packets}
\label{s4-3-5}
Next, we investigated the impact of different numbers of training packets on the mean localization error. We constructed five sets of training and testing datasets with varying numbers of packets for each scenario. From each packet, we extracted 90 average amplitudes, 90 estimated AoAs, and 90 phases from 90 subcarriers of the three antennas. These values were arranged as columns in the CSI images, with the image size being $90 \times W$, assuming $W$ as the number of packets.

Fig.~\ref{fig13} shows that in the lab scenario, the localization error decreases as the number of packets increases. In the corridor scenario, the localization error initially increases and then decreases with the increase in the number of packets. This behavior may be attributed to signal disturbances during certain time intervals. When the number of training packets is 90, both scenarios exhibit the minimum mean localization error, with values of 2.691m (lab) and 2.192m (corridor). This indicates that more diverse information contributes to improved accuracy, and thus, we chose to use 90 training packets in the other experiments for higher precision. Additionally, the overall change in localization error is minimal as the number of packets varies, suggesting the robustness of Secci to the number of training packets.

\begin{figure}[!ht]
	\setlength{\abovecaptionskip}{-0.1cm}   
	\centering
	\includegraphics[width=0.42\textwidth]{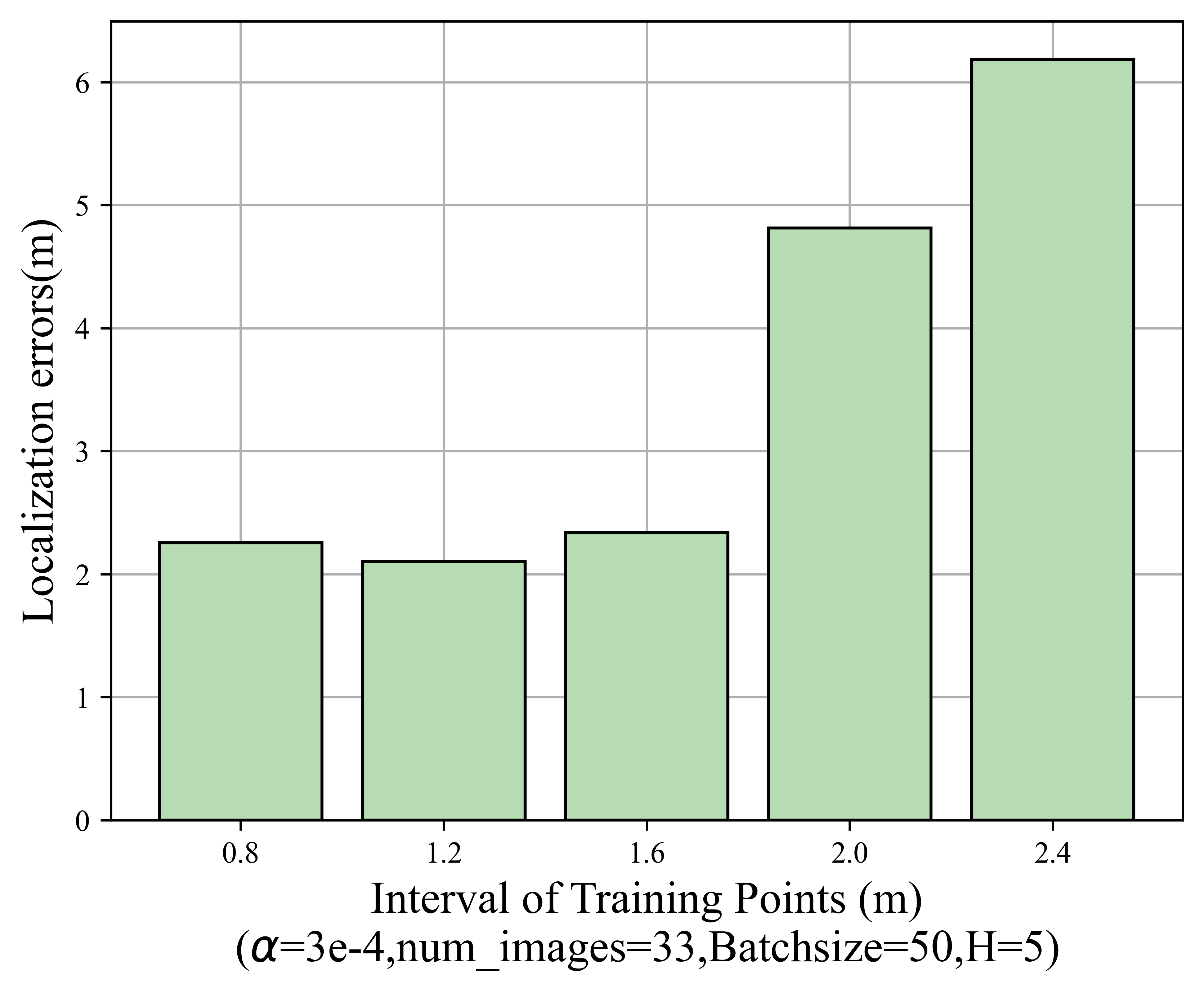}
	\caption{Mean localization errors for different interval between adjacent training points in the corridor scenario.}
	\label{fig14}
\end{figure}
\subsubsection{Effect of interval of Training Points}
\label{s4-3-6}
Lastly, we examined the influence of different intervals between training locations on localization error. Five training and testing datasets were constructed from the corridor environment. Due to environmental constraints, each dataset consisted of a varying number of training locations. The datasets with intervals of 0.8 m and 1.2 m both contained 13 training locations, the dataset with a 2.0 m interval included 10 training locations, and the dataset with a 2.4 m interval included 9 training locations. All testing data were collected near the corresponding training locations. To ensure fairness, all datasets were processed with the same parameter settings in Secci.

Fig.~\ref{fig14} reveals that as the gap between training locations gradually increases, the average localization error also increases. However, there is little difference in localization performance when the gaps are 0.8 m, 1.2 m, and 1.6 m. The minimum error of 2.101 m is achieved with a gap of 1.2 m, while the maximum error of 6.183 m occurs with a gap of 2.4 m. It can be observed that the gap between training locations significantly affects localization performance. If the selected training locations are too sparse, it may lead to fuzziness during the testing phase, resulting in lower localization accuracy. When the gap is small, the training data contains richer positional information, thereby enhancing the prediction performance of Secci. On the other hand, choosing dense training locations would require significant effort in collecting pretraining data, thus presenting a trade-off between cost and prediction accuracy.

\section{Related work}
\label{Related work}
Indoor localization technologies have become increasingly important due to the thriving information industry and the growing demand for location-based services by users\cite{zhu2020indoor}.\,In this section, we discuss two types of indoor localization techniques: intelligent fingerprinting-based\cite{wang2022robust, ren2022fast, wang2019fast} and Angle of Arrival (AoA)-based approaches.

Intelligent fingerprinting-based localization techniques typically utilize five types of fingerprints: RSSI\cite{li2021transloc}, CSI\cite{zhu2021bls}, Bluetooth\cite{lee2018indoor}, magnetic fields\cite{shao2018indoor}, and visible light\cite{guo2017indoor}. To achieve high accuracy in indoor positioning, as mentioned earlier, researchers have proposed six common localization technologies, such as WiFi and RFID. Among them, WiFi has become the primary means of network deployment. WiFi fingerprinting-based localization, which combines signal collection and positioning, has emerged as an effective solution for indoor localization. RSSI is often the preferred choice as a fingerprint due to its simplicity and low hardware requirements. The Horus system utilizes a probabilistic method with RSSI for target location estimation\cite{youssef2005horus}. Research has shown that indoor positioning accuracy based on CSI is significantly higher than that based on RSSI\cite{shi2018accurate}. AF-DCGAN\cite{li2019af} converts CSI into amplitude feature maps and expands the fingerprint database using DCGAN, thereby reducing the manual labor required in the offline phase. CSI, as a promising localization feature, has gradually been adopted as an enhanced channel metric in indoor positioning schemes. Indeed, there have been works that consider both RSS and CSI\cite{li2022long}, proposing a framework called LSTP for Dynamic Fingerprinting.

Next, we discuss the evolution of AoA estimation algorithms. Initially, traditional AoA estimation algorithms can be classified into three main categories: feature-based algorithms\cite{schmidt1986multiple}, probability-based estimation algorithms\cite{ziskind1988maximum}, and sparse representation-based algorithms\cite{yin2011direction}. To address the coherence between multipath signals in indoor environments, spatial smoothing algorithms have been applied\cite{shan1985spatial}. Furthermore, frequency domain-based AoA estimation algorithms have been utilized in OFDM systems, where phase differences can be captured not only between adjacent antennas but also at the subcarrier level\cite{han2017new}. Lastly, spatial domain-based AoA estimation algorithms leverage spatial resources in addition to frequency resources. For instance, in the D-MUSIC system\cite{qian2017enabling}, the device rotates a certain angle around a vertex, and the angles between devices are used to estimate the arrival angles of targets. While AoA-based methods offer high accuracy, their computational complexity renders them unsuitable for real-time applications.

\section{CONCLUSIONS}
\label{CONCLUSION}
This paper introduces Secci, a localization scheme that utilizes CSI images and an attention mechanism-assisted DCNN network. We validate the feasibility of using average amplitude values, estimated AoA values, and phase values as fingerprints through theoretical analysis and experiments. The Secci system architecture and principles are described, utilizing diverse data to construct RGB CSI images for network training and estimating the mobile device's position using newly collected test data. Comprehensive experiments demonstrate the outstanding performance of the Secci system.



 
%

\bibliographystyle{IEEEtran}
\bibliography{myrefarxiv}


%
%
%
%
%
%
%
%
%
%

\end{document}